\documentclass[12pt]{spieman}  
\usepackage{amsmath,amsfonts,amssymb}
\usepackage{graphicx}
\usepackage{setspace}
\usepackage{tocloft}
\usepackage{orcidlink}
\usepackage{xcolor}

\usepackage{subcaption}

\newcommand{\edit}[1]{\textcolor{black}{#1}}
\usepackage{xurl}

\title{The All-sky Medium Energy Gamma-ray Observatory eXplorer (AMEGO-X) Mission Concept}

\author[1,*]{Regina Caputo\orcidlink{0000-0002-9280-836X}}
\author[2]{Marco Ajello\orcidlink{0000-0002-6584-1703}}
\author[1]{Carolyn A. Kierans\orcidlink{0000-0001-6677-914X}}
\author[1]{Jeremy S. Perkins}  
\author[1]{Judith L. Racusin}
\author[35]{Luca Baldini}
\author[3]{Matthew G. Baring\orcidlink{0000-0003-4433-1365}} 
\author[4,21]{Elisabetta Bissaldi\orcidlink{0000-0001-9935-8106}} 
\author[5]{Eric Burns} 
\author[1,6,7]{Nicholas Cannady}
\author[8]{Eric Charles} 
\author[9]{Rui M. Curado da Silva} 
\author[10]{Ke Fang\orcidlink{0000-0002-5387-8138}} 
\author[11,1,6]{Henrike Fleischhack\orcidlink{0000-0002-0794-8780}} 
\author[36]{Chris Fryer\orcidlink{0000-0003-2624-0056}}
\author[31]{Yasushi Fukazawa}
\author[12]{J. Eric Grove} 
\author[2]{Dieter Hartmann} 
\author[13]{Eric J. Howell} 
\author[14]{Manoj Jadhav} 
\author[2]{Christopher M. Karwin\orcidlink{0000-0002-6774-3111}} 
\author[15]{Daniel Kocevski}
\author[40]{Naoko Kurahashi}
\author[39]{Luca Latronico}
\author[16]{Tiffany R. Lewis\orcidlink{0000-0002-9854-1432}}
\author[17]{Richard Leys}
\author[18]{Amy Lien} 
\author[19]{Lea Marcotulli}
\author[1,6,20]{Israel Martinez-Castellanos} 
\author[21]{Mario Nicola Mazziotta\orcidlink{0000-0001-9325-4672}}
\author[1]{Julie McEnery}
\author[14]{Jessica Metcalfe}
\author[22,23]{Kohta Murase}
\author[1,6,7]{Michela Negro\orcidlink{0000-0002-6548-5622}}
\author[24]{Lucas Parker}
\author[12]{Bernard Phlips}
\author[26]{Chanda Prescod-Weinstein}
\author[27,38]{Soebur Razzaque}
\author[20]{Peter S. Shawhan}
\author[2]{Yong Sheng}
\author[28,29]{Tom A. Shutt}
\author[30]{Daniel Shy} 
\author[12]{Clio Sleator}
\author[16]{Amanda L. Steinhebel\orcidlink{0000-0002-4080-2919}}
\author[17]{Nicolas Striebig}
\author[31]{Yusuke Suda}
\author[32]{Donggeun Tak}
\author[37]{Hiro Tajima}
\author[1,6,7]{Janeth Valverde}
\author[1]{Tonia M. Venters\orcidlink{0000-0002-4188-627X}}
\author[1,6,33]{Zorawar Wadiasingh}
\author[12]{Richard S. Woolf\orcidlink{0000-0003-4859-1711}}
\author[12]{Eric A. Wulf}
\author[16]{Haocheng Zhang\orcidlink{0000-0001-9826-1759 }}
\author[34]{Andreas Zoglauer}

\affil[1]{NASA Goddard Space Flight Center, Greenbelt, MD, USA}
\affil[2]{Department of Physics and Astronomy, Clemson University, Clemson, SC 29634, USA}
\affil[3]{Department of Physics and Astronomy, Rice University, 6100 Main Street, Houston, Texas 77251-1892, USA}
\affil[4]{Dipartimento Interateneo di Fisica dell'Universit\'{a} e del Politecnico di Bari, via Amendola 173, 70126, Bari, Italy}
\affil[5]{Department of Physics \& Astronomy, Louisiana State University, Baton Rouge, LA 70803, USA}
\affil[6]{Center for Research and Exploration in Space Science and Technology, NASA/GSFC, Greenbelt, Maryland 20771, USA}
\affil[7]{University of Maryland, Baltimore County, Baltimore, MD 21250, USA}
\affil[8]{W. W. Hansen Experimental Physics Laboratory, Kavli Institute for Particle Astrophysics and Cosmology, Department of Physics and SLAC National Accelerator Laboratory, Stanford University, Stanford, CA 94305, USA}
\affil[9]{Laborat\'{o}rio de Instrumenta\c{c}\~{a}o e F\'{i}sica Experimental de Part\'{i}culas, Departamento de F\'{i}sica, Universidade de Coimbra, P-3004-516 Coimbra, Portugal}
\affil[10]{Department of Physics, University of Wisconsin-Madison, Madison, Wisconsin, USA}
\affil[11]{Catholic University of America, 620 Michigan Ave NE, Washington, DC 20064, USA}
\affil[12]{Space Science Division, U.S. Naval Research Laboratory, Washington, DC, 20375, USA.}
\affil[13]{OzGrav, University of Western Australia, Crawley, Western Australia 6009, Australia}
\affil[14]{Argonne National Laboratory, Lemont, IL 60440, USA}
\affil[15]{NASA Marshall Space Flight Center, Huntsville, AL 35808, USA}
\affil[16]{NASA Postdoctoral Fellow, NASA Goddard Space Flight Center, Greenbelt, MD, USA}
\affil[17]{Karlsruhe Institute of Technology (KIT-ADL) - Hermann-von-Helmholtz-Platz 1, D-76344 Eggenstein-Leopoldshafen}
\affil[18]{University of Tampa, Department of Chemistry, Biochemistry, and Physics, 401 W. Kennedy Blvd, Tampa, FL 33606, USA}
\affil[19]{Department of Physics, Yale University, 52 Hillhouse Avenue, New Haven, CT 06511, USA}
\affil[20]{Department of Physics, University of Maryland, College Park, Maryland 20742, USA}
\affil[21]{Istituto Nazionale di Fisica Nucleare, Sezione di Bari, Via E. Orabona 4, 70125 Bari, Italy}
\affil[22]{Dept. of Physics and Dept. of Astronomy and Astrophysics, Institute for Gravitation and the Cosmos, The Pennsylvania State University, University Park, Pennsylvania, USA}
\affil[23]{Center for Gravitational Physics, Yukawa Institute for Theoretical Physics, Kyoto, Japan}
\affil[24]{Los Alamos National Laboratory, Los Alamos, NM 87544, USA}
\affil[25]{Space Science Division, U. S. Naval Research Laboratory, Washington, DC 20375, USA}
\affil[26]{Department of Physics \& Astronomy, University of New Hampshire, Durham, NH 03824, USA }
\affil[27]{Centre for Astro-Particle Physics and Department of Physics, University of Johannesburg, PO Box 524, Auckland Park 2006, South Africa}
\affil[28]{SLAC National Accelerator Laboratory, Menlo Park, CA 94025, USA }
\affil[29]{Kavli Institute for Particle Astrophysics and Cosmology, Stanford University, Stanford, CA 94305, USA}
\affil[30]{National Research Council Research Associate resident at the Naval Research Laboratory, Washington DC, 20375, USA}
\affil[31]{Department of Physics, Hiroshima University, 1-3-1 Kagamiyama, Higashi-Hiroshima, Hiroshima, Japan, 739-8526}
\affil[32]{Deutsches Elektronen-Synchrotron (DESY), Platanenallee 6, Zeuthen, 15738, Germany}
\affil[33]{Department of Astronomy, University of Maryland, College Park, Maryland 20742, USA}
\affil[34]{Space Sciences Laboratory, University of California at Berkeley, 7 Gauss Way, Berkeley, CA 94720, USA}
\affil[35]{Universit\'{a} di Pisa and Istituto Nazionale di Fisica Nucleare, Sezione di Pisa I-56127 Pisa, Italy}
\affil[36]{Center for Theoretical Astrophysics, Los Alamos National Laboratory, Los Alamos, New Mexico, 87545, USA}
\affil[37]{Institute for Space–Earth Environmental Research and Kobayashi-Maskawa Institute for the Origin of Particles and the Universe, Nagoya University, Japan}
\affil[38]{Department of Physics, The George Washington University, Washington, DC 20052, USA}
\affil[39]{INFN, Sezione di Torino, Via Pietro Giuria 1, 10125 Torino, Italy}
\affil[40]{Department of Physics, Drexel University, Philadelphia, PA, USA, 19104}

\cftpagenumbersoff{figure}
\cftpagenumbersoff{table} 
\begin{document} 
\maketitle

\begin{abstract}

The All-sky Medium Energy Gamma-ray Observatory eXplorer (AMEGO-X) is designed to identify and characterize gamma rays from extreme explosions and accelerators. 
The main science themes include: supermassive black holes and their connections to neutrinos and cosmic rays; binary neutron star mergers and the relativistic jets they produce; cosmic ray particle acceleration sources including Galactic supernovae; and continuous monitoring of other astrophysical events and sources over the full sky in this important energy range. 
AMEGO-X will probe the medium energy gamma-ray band using a single instrument with sensitivity up to an order of magnitude greater than previous telescopes in the energy range 100 keV to 1 GeV that can be only realized in space. 
During its three-year baseline mission, AMEGO-X will observe nearly the entire sky every two orbits, building up a sensitive all-sky map of gamma-ray sources and emission.
AMEGO-X was submitted in the recent 2021 NASA MIDEX Announcement of Opportunity.

\end{abstract}


{\noindent \footnotesize\textbf{*}Corresponding author  \linkable{regina.caputo@nasa.gov} }

\begin{spacing}{1}   

\section{Introduction}
Multimessenger astrophysics is one of  the most exciting and rapidly advancing fields of science, providing unparalleled access to extreme processes that sculpt the universe.  Gamma-ray observations have been central to the advent of the multimessenger initiative and will continue to be so as the field matures. 
As a priority theme of the Astro2020 Decadal Survey report~\footnote{\url{https://www.nationalacademies.org/our-work/decadal-survey-on-astronomy-and-astrophysics-2020-astro2020}}, the science of New Messengers New Physics is poised to revolutionize our understanding of the extreme universe. 
Data from NASA’s {\it Fermi} mission has demonstrated that many of the extreme processes that produce gravitational waves and neutrinos and accelerate cosmic rays also produce gamma rays. 
In other words, \textit{multimessenger sources are gamma-ray sources}. 
These sources are, however, brightest in the under-explored MeV band.
We have developed a mission and submitted a proposal to the 2021 NASA Astrophysics Medium Explorer (MIDEX) Announcement of Opportunity~\footnote{\url{https://explorers.larc.nasa.gov/2021APMIDEX/MIDEX/index.html}} to observe these critical energies and fully capitalize on this exciting new era of multimessenger astrophysics.

The All-sky Medium Energy Gamma-ray Observatory eXplorer (AMEGO-X) is a wide-field survey telescope designed to discover and characterize gamma-ray emission from multimessenger sources using imaging, spectroscopy, and polarization.
During its three-year baseline mission, AMEGO-X will observe in the critical 100 keV - 1 GeV energy band over nearly the entire sky every two orbits, building up a sensitive all-sky map of gamma-ray sources and diffuse emission. 
It will also access $>$50\% ($<$10 MeV) and $>$20\% ($>$10 MeV) of the sky instantaneously, maximizing transient detections and rapid alerts, which will be openly distributed to the astrophysics communities. 
AMEGO-X will deliver breakthrough discoveries \edit{in a} MIDEX \edit{class mission} in areas highlighted as the highest scientific priority for Explorer-scale missions in the Astro2020 Decadal Survey Report: multimessenger astrophysics and time-domain astronomy. 
AMEGO-X complements the recently selected COSI Small Explorer mission~\cite{Tomsick:2021wed}, which has excellent energy resolution in the 0.2--5\,MeV band to probe the origins of Galactic positrons and uncover sites of nucleosynthesis. 
AMEGO-X will provide a leap ($\times$10--50) in continuum sensitivity, similar to Fermi/LAT at higher energies, in the long sought-after MeV gamma-ray gap.
The AMEGO-X mission employs a single instrument with subsystems delivered by Argonne National Laboratory, the Naval Research Laboratory and NASA Goddard Space Flight Center and has partnered with Lockheed Martin Space for the high-heritage spacecraft.

\section{Science}
After a long hiatus from the first multimessenger event SN 1987A~\cite{sn1987a}, which was discovered in optical and neutrinos, a new era of multimessenger astrophysics was heralded in 2017 when the electromagnetic counterparts to sources of both neutrinos~\cite{science.aat2890, science.aat1378} and gravitational waves~\cite{LIGOScientific:2017vwq, LIGOScientific:2017ync} were observed for the first time. 
The AMEGO-X mission has three overarching goals which tie together the extreme explosions and extreme accelerators which produce all the cosmic messengers: supermassive black holes, neutron stars and their mergers, and the remnants of Galactic supernovae. 
The following section will expand upon these overarching science goals. 

\subsection{Supermassive Black Holes and their connections to Neutrinos and Cosmic Rays}
Active galactic nuclei (AGN), accreting supermassive black holes at the centers of galaxies, provide extreme conditions that are conducive to the particle acceleration that is inferred from observations of non-thermal X rays and gamma rays.
This can take place near the black hole (core) and/or in the relativistic jets that 10\% of AGN display. 
Those AGN with a relativistic jet aligned along our line of sight are called blazars~\cite{Urry_1995} and are the most powerful persistent sources of electromagnetic radiation in the universe. 
Blazars are widely believed to be sites for rapid acceleration of electrons in their jets to near the speed of light  because their low energy spectra are consistent with electron synchrotron emission and optical polarization measurements confirm this~\cite{Boettcher:2013wxa}. 
They are also powerful enough to accelerate protons and may be responsible for the generation of ultra-high energy cosmic rays (UHECRs). 
The detection of the high-energy neutrino, IC-170922A, by the IceCube Neutrino Observatory (IceCube) coincident with the brightly flaring gamma-ray blazar, TXS 0506+056, marked a milestone in multimessenger astrophysics~\cite{science.aat2890, science.aat1378}. 
It established observationally that at least some blazars accelerate protons: neutrinos are a unique signature of proton acceleration and production because they cannot be produced in purely leptonic sources, instead being produced as decay products of pions and muons in photohadronic collisions. 
However, the picture is far from complete, as it is not clear which among jetted or non-jetted AGN can accelerate protons. 
The ``smoking gun" signatures of proton acceleration in AGN are either the coincident detection of high-energy PeV neutrinos and medium-energy (MeV) gamma rays in a particular AGN, and/or detection of polarized emission~\cite{Zhang:2013bna} in the MeV band. 
\edit{AMEGO-X, with its sensitive all-sky monitoring, will detect over 400 blazars, which is determined considering the log N - log S distributions of Swift/BAT and Fermi/LAT blazars~\cite{2009ApJ...699..603A, 10.1093/pasj/psv043, Tsuji:2021zfc}. 
Using the 2nd FAVA catalog~\cite{2017ApJ...846...34A}, it will also detect around 150 blazar flares/year, and for the brightest of these also measure the polarization fraction for more than 10 blazar flares/year, which is the sensitivity to measure polarization within a week.}
This will allow us to answer long-standing questions about the source of cosmic neutrinos, extragalactic cosmic rays, while also providing critical gamma-ray capabilities complementary to upgraded neutrino observatories in the era of multimessenger astrophysics.

\subsubsection{Blazars and their Flares}
\begin{figure}
\begin{center}
\includegraphics[width=1.0\textwidth]{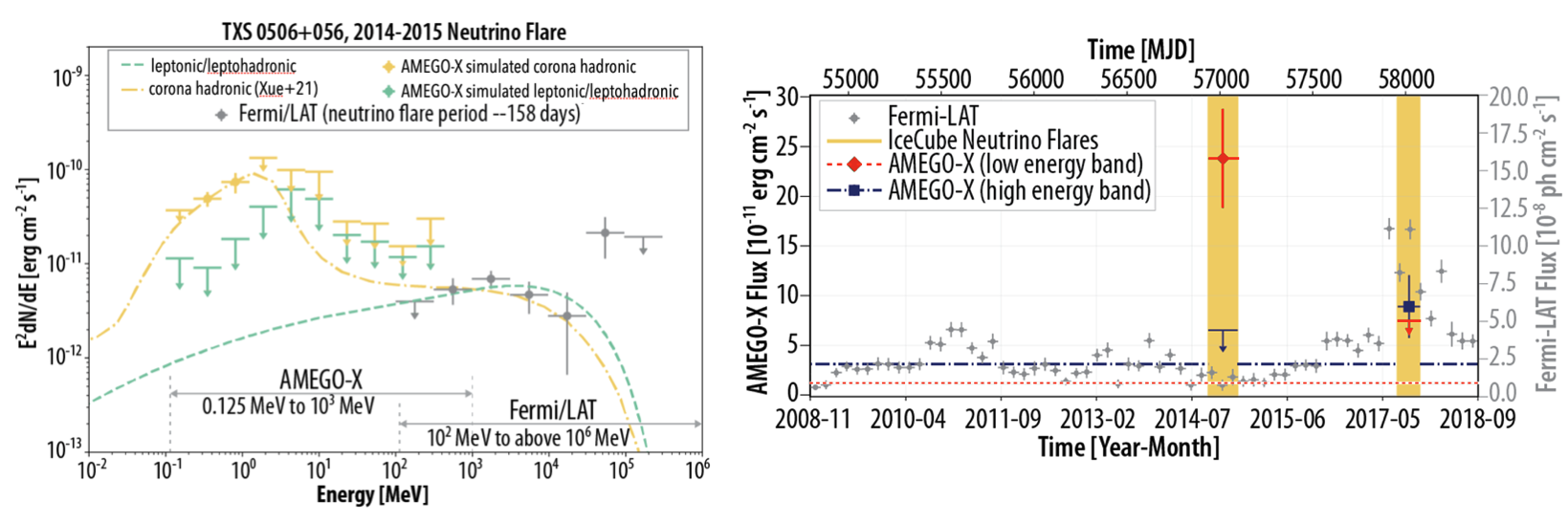} 
\end{center}
        \caption{Simulated AMEGO-X SED for the 2014 flare of TXS~0506+056. The green dashed line is representative of leptonic and leptohadronic models, while the yellow dot-dashed line is the two-zone hadronic model where one of the zones is from the AGN corona from Ref. \cite{xue2021}. The gray data points are measurements from Fermi/LAT (left). Light-curve of TXS~0506+056. The gray data points are measurements from Fermi/LAT. The yellow band show the times of the IceCube neutrino flares (right). The green data points are simulated AMEGO-X detection of the MeV emission during those flares. See \cite{Lewis:2021roc} for more details.
        }
        \label{fig:d1}
\end{figure}

Relativistic jets produce gamma rays through the interactions of the particles they accelerate. 
In leptonic models, accelerated electrons produce all observed gamma rays, principally via synchrotron and inverse Compton emission.
In hadronic models, accelerated protons produce gamma rays through proton-synchrotron radiation~\cite{Mucke:2000rn} or via synchrotron radiation from secondary particles generated by proton interactions with jet photons (via e.g., photo-pion production), the latter of which also produces neutrinos~\cite{Mannheim:1993jg}. 
For most flare data from currently operating observatories, it is not possible to determine which scenario better describes the data – both hadronic and leptonic models are consistent with observed blazar spectra~\cite{Boettcher:2013wxa, Keivani:2018rnh, Reimer_2020}. 
Some hadronic emission dominated models predict that significant high-energy (GeV) gamma rays interact with jet photons (effectively being absorbed) and are reprocessed into the medium-energy gamma-ray band where they can finally escape the jet. 
By monitoring the entire medium-energy gamma-ray (100 keV-1GeV) sky every \edit{three} hours, AMEGO-X will observe in real time the potential sources of IceCube neutrinos and correlate the timing between gamma-ray and neutrino flares. 
Figure~\ref{fig:d1} (left) shows a simulated AMEGO-X light curve for which the first and second neutrino flare of TXS 0506+056 could have been detected, since gamma rays and neutrinos are produced via the same processes~\cite{Lewis:2021roc}.
While hadronic models link neutrinos with gamma rays in the AMEGO-X band, the same cannot be said for higher energies. 
The brightest neutrino events may not be detected by higher-energy gamma-ray telescopes, such as the Fermi/LAT, because the radiation fields required for efficient neutrino production make the source opaque to high-energy gamma rays [e.g., \cite{Murase:2015xka, Reimer_2020}]. 
Figure~\ref{fig:d1} (right) shows that during the 2014-2015 TXS 0506+056 orphan neutrino flare, AMEGO-X may have detected a significant MeV flare if the emission was from hadronic interactions within the AGN corona.

By monitoring nearly the full sky every three hours, AMEGO-X will detect $\sim$150 blazar flares/ year, the most promising counterparts to IceCube high-energy neutrinos \edit{which are} released through real-time alerts~\cite{IceCube:2016cqr}. 
AMEGO-X will provide an all-sky catalog of more than 400 MeV-peaked blazars for neutrino searches~\cite{2009ApJ...699..603A}. 
These are the most powerful blazars, whose maximum energy output is in the AMEGO-X band. 
{In blazars, the number of high-energy neutrinos correlates with the total 1 keV - 1 PeV flux, which is well approximated by the 100 keV - 1 GeV flux, since this is the band where blazars release the most energy~\cite{krauss2018}. 
The number of expected IceCube-AMEGO-X coincident detections (over three years) for the baseline and threshold missions are 7.7($\pm$2.5) and 6($\pm$2), respectively (See Section~\ref{Section:Instrument}), with the brightest sources expected to emit more than one high-energy neutrino within the 3 years. 
This has been computed from the number of expected neutrinos from each source following \cite{krauss2018}:
\begin{equation}
    N_{\nu} = N_{\nu,max}(f_{blazar}) * f_{corr} * 3/4
\end{equation}
where $N_{\nu,max}$ is the maximum number of expected neutrinos (in 4 years) for a given blazar with \edit{an average} flux $f_{blazar}$.   $f_{corr}$=0.0089 is a correction that takes into account realistic blazar spectra, neutrino flavors and different classes of blazars (conservative estimate~\cite{krauss2018}).
The factor 3/4 accounts for the different integration time (4 years in Ref.~\cite{krauss2018}) compared with the expected 3-year duration of the AMEGO-X survey). 
}
Moreover, the AMEGO-X survey of the most powerful blazars will also be crucial to determine their contribution to the IceCube diffuse astrophysical neutrino flux, the origins of which remain a mystery~\cite{IceCube:2019cia}.

\subsubsection{Polarization measurements of Jetted Active Galactic Nuclei}

\begin{figure}
	\begin{center}
	\includegraphics[width=1.0\textwidth]{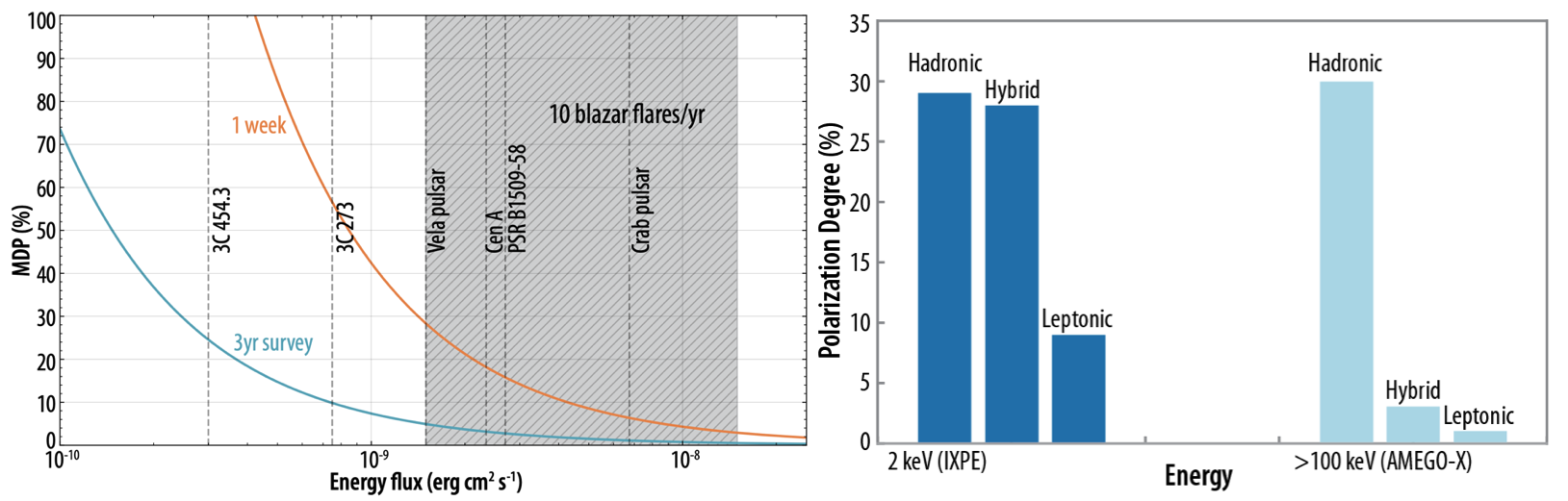} 
	\end{center}
        \caption{With a minimum detectable polarization of 25\,\% \edit{for faint sources during the 3 year survey (left)}, AMEGO-X will be able to discover blazars whose high-energy emission is of hadronic nature\cite{Zhang:2013bna}.
        \edit{(Right) AMEGO-X observations complement IXPE's because polarization at  different energies probes different blazar populations, radiation processes and possibly even different acceleration mechanisms.} Polarization fractions for blazars with a fully hadronic, leptonic or hybrid (mix of leptonic and hadronic) emission models in the IXPE and AMEGO-X band. 
        }
        \label{fig:d1_pol_ixpe}
\end{figure}

Multiwavelength spectral observations alone cannot yet distinguish if blazars accelerate only leptons or leptons and hadrons [e.g., ~\cite{Cerruti:2018zxo, Murase:2018iyl}]. 
In blazar models for which at least part of the gamma-ray emission is from proton acceleration, the polarization signature is prominent in the medium-energy gamma-ray band~\cite{Zhang:2013bna}. 
This is because the relativistic Compton scattering dominates the high-energy emission in leptonic models and produces a much lower fraction of polarization than the proton and cascade synchrotron emission in the hadronic model~\cite{Paliya:2018wgu}. 
AMEGO-X polarization measurements of blazars will provide an independent constraint on proton acceleration in AGN complementary to coincident neutrino detection~\cite{Zhang:2019dob}. 
While IXPE \cite{ixpe} can detect X-ray polarization from blazar jets,  \edit{this polarization probes the acceleration mechanism of the relativistic electrons, which produce the emission in the X-ray band.} On the other hand,
AMEGO-X will explore hadronic signatures via MeV polarimetry for blazars, unambiguously identifying proton acceleration and neutrino production in blazar jets, which are essential to multimessenger astronomy. 
Blazar spectral models have shown that many BL Lacertae objects (BL Lacs), in particular intermediate-synchrotron-peaked and high-synchrotron-peaked BL Lacs, have a significant or even dominating contribution from the primary electron synchrotron component~\cite{peirson2022}. 
In these cases, IXPE detected polarization would \edit{indicate leptonic, not hadronic, interactions} in the X-ray band. 
Furthermore, MeV polarimetry can uniquely probe proton synchrotron emission in blazar jets. 
As shown in Figure~\ref{fig:d1_pol_ixpe}, AMEGO-X can clearly distinguish the model with leptonic Compton scattering and hadronic cascades (hybrid model) versus the fully hadronic models with proton synchrotron emission  (hadronic model), which is not possible at lower energies. \edit{This is because these sources, mostly flat-spectrum radio quasars, are too faint at X-ray energies for IXPE to distinguish polarization degrees of a few percentage (like those, e.g., in Figure~\ref{fig:d1_pol_ixpe}.)}
Discovering fully hadronic sources (those with proton synchrotron emission) will allow AMEGO-X to unveil the origin of ultra-high-energy cosmic rays, something not possible without observations in the MeV gamma-ray regime.

AMEGO-X will measure the gamma-ray polarization fraction as low as 10\% for at least 10 bright blazar flares per year (e.g., from 3C 279, PKS 1510-089) and two nearby persistent blazars (3C 273, 3C 454.3), determining whether AGN jets accelerate high-energy cosmic-ray (CR) protons for the first time at these energies. 
A measured polarization fraction $>$30\% will allow AMEGO-X to independently establish blazars as extragalactic CR accelerators ~\cite{Zhang:2013bna, Paliya:2018wgu}.

\subsubsection{The AGN Core}

\begin{figure}[ht]
\begin{center}
\includegraphics[width=0.50\textwidth]{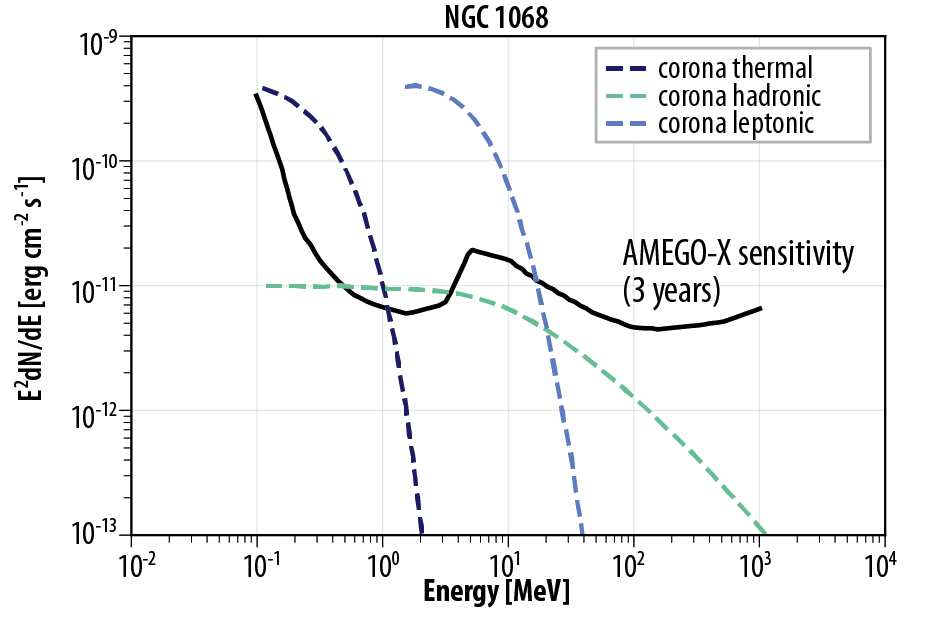} 
\end{center}
\caption{Expected emission from the AGN corona of NGC~1068. The blue dashed line shows the corona thermal emission, the light blue and green dashed lines show the non-thermal leptonic and hadronic components respectively \cite{Murase:2019vdl}. \edit{The solid black line shows the AMEGO-X sensitivity assuming a 3 year mission.}}
\label{fig:d2}
\end{figure}

IceCube has recently identified NGC 1068, a nearby AGN lacking a relativistic jet, as a potential neutrino source~\cite{IceCube:2020acn}; this provides evidence that AGN without a jet may also accelerate protons and produce neutrinos. 
Within the immediate vicinity of the supermassive black hole (AGN core), accretion cultivates extreme environments conducive to particle acceleration. 
Clouds of hot thermal electrons exist above the disk in the corona and have a thermal energy distribution~\cite{2001MNRAS.321..549M}. 
However, there is recent evidence of accelerated (non-thermal) electrons that may produce MeV emission via inverse Compton scattering of disk photons~\cite{2018ApJ...869..114I, Inoue:2019fil}. 
Similarly, protons can be accelerated near the disk~\cite{Murase:2019vdl}. 
AMEGO-X’s continuum sensitivity will allow detection of these thermal and non-thermal leptonic and/or hadronic components (Figure~\ref{fig:d2} for NGC 1068) in off-axis jetted (e.g., Cen A, M87) and non-jetted (e.g., NGC 1068, NGC 3516, NGC 4258) AGN.

\subsection{Binary Neutron Star Mergers and their Jets}

In-spiraling and merging binary neutron stars (BNSs) emit gravitational waves (GWs), ripples in spacetime that can be detected by the worldwide GW network~\cite{gw_network}. 
After the merger, particles are accelerated in narrowly collimated relativistic jets launched along the rotational axis. 
When these jets point towards Earth, we observe a bright flash of gamma rays, a short gamma-ray burst (SGRB), in the tens of keV to few MeV energy band. 
GW170817/GRB170817A was the first BNS merger jointly detected both with gamma rays and GWs, informing telescopes around the world to follow up the event~\cite{LIGOScientific:2017vwq, LIGOScientific:2017ync, Abbott_2017_emgw}. 
Despite the wealth of information gained through multi-wavelength observations of this event and the associated radioactively powered isotropic kilonova emission~\cite{Metzger:2017wot} and the off-axis afterglow, fundamental questions regarding the nature of BNS mergers remain. 
For example, a higher-mass BNS merger, GW 190425\cite{gw190424}, had no associated SGRB detected by current missions, which could be due to a different merger remnant or to an off-axis jet that was pointed away from Earth. \edit{The large 90\% localization region of 10,200 deg$^2$ was not fully covered by current instruments. 
Swift/BAT reported that 45.73\% of the localization probability region was outside the field-of-view~\cite{2019GCN.24184....1S} and only 55.6\% of the probability region was viewable by Fermi/GBM~\cite{2019GCN.24185....1F}}. 
A population of joint observations of BNS mergers, in GW and gamma rays, will address these questions, enabling science which is impossible to perform independently or at other wavelengths. 
The medium-energy gamma-ray band, where SGRBs are the brightest, provides unique insight into the physics of BNS mergers making AMEGO-X ideally positioned to provide a new, comprehensive view of this population.

\subsubsection{Model Dependent Statistical Fractions and Structure of GRB jets}

\begin{figure}
\begin{center}
\includegraphics[width=1.0\textwidth]{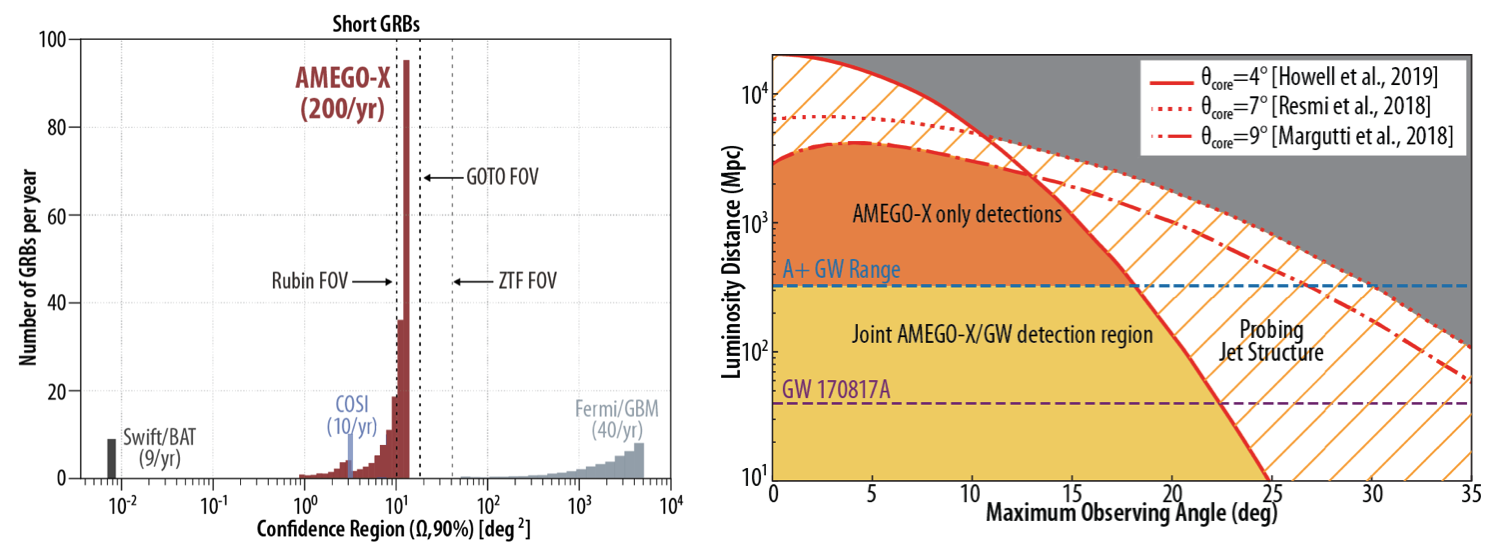} 
\end{center}
        \caption{Comparison of the detection rate vs localization accuracy for AMEGO-X, Swift/BAT, Fermi/GBM and COSI (left). The distance by AMEGO-X, versus jet observing angle for SGRBs (right). The solid, dot-dashed and dotted lines show jet structure models (from respectively Refs.~\cite{2019MNRAS.485.1435H,resmi2018,margutti2018}) compatible with the observations of the GRB170817A. An event falling within the yellow shaded area will be detected both by AMEGO-X and the GW network. Observations or non-observations of events that lie below the blue dashed line and within the hatched area will indicate the extent of the GRB jet opening angle. Events in the orange area are detectable by AMEGO-X, but outside of the range of the GW network. }
        \label{fig:d3andd4}
\end{figure}

A BNS merger can produce one of four possible remnants: a stable NS, a long-lived massive NS supported by its fast rigid-body like rotation, a short-lived massive NS supported by internal differential rotation, or prompt black hole formation. 
Each outcome leads to different predictions for the presence or absence of a relativistic jet, as well as the characteristic delay between the merger event and the formation and launching of the jet~\cite{Burns:2019byj}. 
For events detected jointly (estimates are later in the section), the GW data provide information on NS masses, spins, and distances, but AMEGO-X observations are required to probe the nature of the merger remnant and the structure of the jet. 
Joint observations of ten BNS mergers can constrain the maximum NS mass to $\sim$1\% precision, improving the current constraints by nearly an order of magnitude~\cite{Margalit:2019dpi}. 
Typical AMEGO-X on-board SGRB localizations will have a factor of 400 smaller localization uncertainties (90\% confidence regions) compared to Fermi/GBM~\cite{Connaughton:2014xha, Goldstein:2019pcj}, which will dramatically decrease the time needed to search for and precisely localize the SGRB afterglow or kilonova emission via X-ray, optical, and radio follow-up.  This increases the fraction of events with early broadband observations including spectroscopy and an opportunity to characterize the early evolution of these objects.
The AMEGO-X $\lesssim$10 deg$^2$ 90\% confidence region (Figure~\ref{fig:d3andd4}, left) is well matched to the Fields of View (FOVs) of current facilities like the Gravitational wave Optical Transient Observer [GOTO ~\cite{Steeghs:2021wcr}] and the Zwicky Transient Facility [ZTF ~\cite{2019PASP..131a8002B}], and the future Vera C. Rubin Telescope~\cite{2019ApJ...873..111I}. 
Those multiwavelength observations will provide insights on the host galaxy including redshift, source dynamics and kilonova emission, complementing the AMEGO-X jet energetics and emission mechanisms. 
AMEGO-X’s rapid SGRB localizations enable the $>$2000 observers (subscribers to the gamma-ray coordinates network, GCN) to quickly search and identify the potential counterparts. 
This population would inform not only the uniqueness of GW170817 but also the nature of the progenitors and remnant objects from compact binary mergers.

Notably under-luminous compared to previously detected SGRBs~\cite{LIGOScientific:2017vwq, Abbott_2017_emgw}, the energetic core of GRB170817A’s relativistic jet was not completely aligned with our line-of-sight~\cite{2018Natur.561..355M} ($\sim$20-25$^{\circ}$ off-axis). 
The observed emission supports the structured jet scenario, in which the jet interacts with the ejecta from the BNS merger resulting in a narrowly collimated energetic jet core with a gradual decrease in flux and Lorentz factor away from the core ~\cite{2017MNRAS.471.1652L, Xie:2018vya}. 
Under this structured jet scenario, the observed luminosity has a strong dependence on the viewing angle with respect to the jet axis and can be parameterized by a Gaussian function defined by the opening angle ($\theta_{\mathrm{core}}$) and luminosity of the jet (Figure~\ref{fig:d3andd4}, right panel) and only nearby SGRBs can be detected at wider viewing angles. 
Most BNS mergers are expected near the GW detectability horizon (or the maxiumum distance the GW network is sensitive) where the volumetric rate will be higher.
This distance-viewing angle phase space (Figure~\ref{fig:d3andd4}, right panel) provides a means to directly probe the structured jet profile. Here we note that both AMEGO-X detection and non-detections of GW-detected BNS mergers will set fundamental constraints on the structure of the jet.

Of the 80$^{+180}_{-50}$/yr GW-detected BNS mergers that will fall within AMEGO-X’s FOV in the late 2020s (all orientations)~\cite{LIGOScientific:2020kqk},$\sim$40\% (30$^{+68}_{-19}$/yr) will have jets oriented within 30$^{\circ}$ \edit{with a 90\% CL}. 
\edit{This assumes 80\% duty cycle per GW detector (ie: 64\% for both LIGO detectors) and that AMEGO-X observes 50\% of the sky instantaneously for GRBs}. 
Among those, the fraction of detected SGRBs will strongly constrain the jet structure. For example, for the narrowest and widest jet models of Figure~\ref{fig:d3andd4} (compatible with GRB170817A) one expects an average $\sim$10\% (3$^{+5}_{-2}$/yr) and $\sim$25\% (6$^{+10}_{-4}$/yr) joint GW-SGRB detections, respectively~\cite{2019MNRAS.485.1435H}. 
Knowledge of the GW distances will allow AMEGO-X observations (both detections and non-detections) to place constraints on different jet structure models and their parameterization such as the parameters of the Gaussian structured jet, or more complicated models such as two-component structured jet (Figure~\ref{fig:d3andd4}) ~\cite{2019MNRAS.485.1435H}. 
Although not the favored scenario for GRB170817A, a cocoon shock breakout can produce SGRBs over a much wider range of viewing angles, despite carrying only a fraction of the energy of the jet~\cite{2017Sci...358.1559K}. 
In all jointly observed GW-SGRB events, AMEGO-X will be able to detect or rule out the soft ($<$200 keV) emission associated with shock breakout. This population of BNS mergers will constrain the jet parameters providing insights into the formation of relativistic outflows.

Beyond BNS mergers, AMEGO-X will also provide observations of more than a hundred GW-detected neutron star black hole (NSBH) mergers during its mission lifetime. 
These observations will determine if NSBH mergers emit gamma rays~\cite{Foucart:2020ats}. 
Without measurements in the medium-energy gamma-ray band, a critical piece for understanding the remnant and potential jet production of the GW-detected NSBH mergers will be missing.

Recent Very High-Energy (VHE) detections of GRB afterglows have accentuated the issue of whether gamma-ray afterglows violate the maximum synchrotron energy or if a synchrotron self Compton component is needed to explain GeV/TeV afterglows~\cite{lat130427a,hess180720b,magic190114c}.  AMEGO-X will measure afterglow spectra between \edit{hard} X-ray and GeV, where the distinction between these models is most significant.  Previously we have only been able to interpolate between \edit{hard} X-ray and GeV energies~\cite{ajello2018}, with the exception of a handful of NuSTAR observations~\cite{kouveliotou2013}.

\subsection{Cosmic ray sources in the Galaxy}

Energetic charged particles (CRs, mostly protons and electrons) are ubiquitous in our galaxy. 
They are accelerated in a variety of environments, such as shocks and regions with strong magnetic fields. Stellar nurseries and remnants of massive stellar explosions provide such environments. 
After decades of observations, two main questions regarding galactic CRs remain: in what environments are protons accelerated and what is the origin of the positron excess? 
AMEGO-X, with its projected sensitivity, and timing accuracy, will open the possibility to discover new galactic proton accelerators and test whether pulsars are the source of the positron excess.

\subsubsection{Supernova remnants, Novae and Star-formation Regions}

\begin{figure}[ht]
\begin{center}
\includegraphics[width=0.75\textwidth]{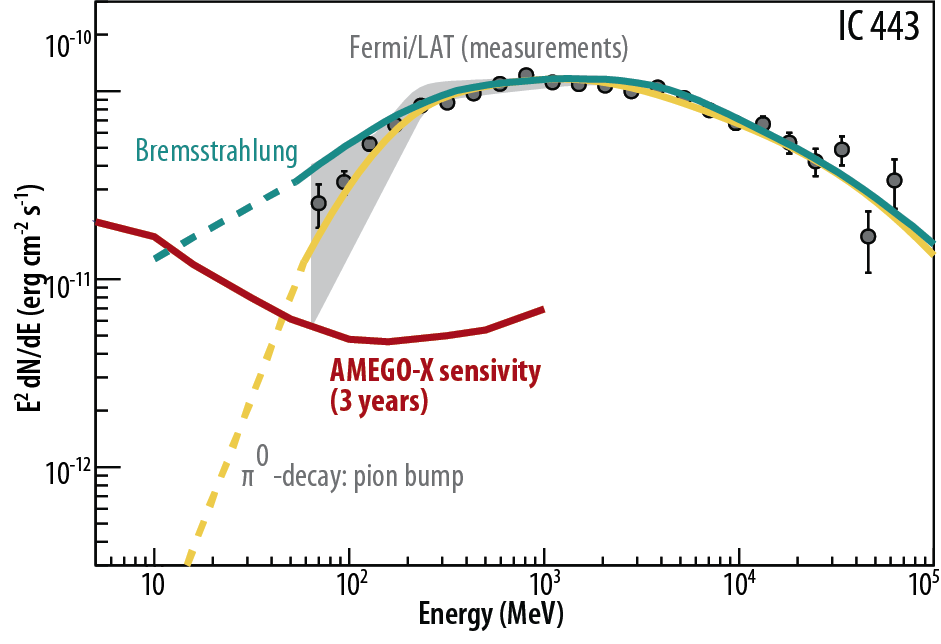} 
\end{center}
\caption{Fermi/LAT SED of SNR IC~443. The green and yellow lines show the Bremsstrahlung (with break) and $\pi^0$ decay models from Ref.\cite{pion_bump}. The red solid line shows the AMEGO-X 3\,$\sigma$ sensitivity.
}
\label{fig:d5}
\end{figure}

The smoking gun to identify accelerators of CR protons is to detect the characteristic neutral pion-decay ($\pi_0\rightarrow 2\gamma$) feature, or pion bump, produced in the interaction of protons with the interstellar material. 
Each photon produced has an energy of 67.5 MeV (in the $\pi_0$ rest frame) ~\cite{1971NASSP.249.....S}, which is ideally matched to the AMEGO-X band. 
Proton acceleration is thought to happen in supernova remnants (SNRs), novae, and star-forming regions (SFRs). 
In SNRs, a strong shock, launched by the supernova, sweeps up the ambient medium and provides an ideal site for proton acceleration~\cite{Caprioli:2011lfx}. 
In novae, the material accreted on the white dwarf from the companion eventually undergoes explosive thermonuclear burning, creating a shock that can accelerate protons in the interaction with the companion’s wind~\cite{Chomiuk:2020zek}. 
In SFRs, supernovae shocks and massive stars’ winds provide the means to accelerate protons and low-energy cosmic rays ($<$1 TeV) play a fundamental role in shaping the chemical richness of the interstellar medium, determining the dynamical evolution of molecular clouds~\cite{2011Sci...334.1103A, 2020SSRv..216...29P}. 
However, even in the best-studied case (e.g., SNR IC 443; Figure~\ref{fig:d5}), the models are ambiguous, within the large observational uncertainties because of the lack of observations below 60 MeV. 
AMEGO-X, with its smaller  PSF (factor $>$2) at 60\,MeV than Fermi/LAT (see Sec.~\ref{sec:performance}), will measure the gamma-ray energy spectra of the most promising galactic hadronic accelerators, beyond SNR IC 443, including W44, and W51C; the star-forming regions Cygnus Cocoon, Westerlund 1 and 2; the recurrent T CrB and RS Oph novae and $\sim$one nova/year~\cite{Chomiuk:2020zek}, and will determine the gamma-ray production mechanisms to confirm potential hadronic accelerators found by Fermi/LAT. 
\edit{Based on extrapolating the best fit spectra from the Fermi/LAT 4FGL DR2 catalog~\cite{Fermi-LAT:2019yla, Ballet:2020hze} and the Fermi/LAT Supernova Remnant catalog~\cite{Fermi-LAT:2015xeq}, AMEGO-X will detect 20-40 other SNRs. }
Even if these sources show no evidence of a pion bump, studying energies below 200 MeV will allow measurements of the Bremmstrahlung emission. 
This, together with radio observations, will allow AMEGO-X to determine the magnetic field\cite{1980MNRAS.191..855C} in these environments and provide insight into particle acceleration in the galaxy.

\subsubsection{Pulsars and Pulsar Wind Nebulae}

The flux of galactic CR positrons (electron anti-particles) in the 10-200 GeV range is well measured, yet it exceeds expectations from positrons generated by propagating CRs in interactions with interstellar gas~\cite{2012PhRvL.108a1103A, 2013PhRvL.110n1102A}. 
These additional positrons, which are referred to as a positron excess, could be produced by the known leptonic accelerators, such as pulsars, or by new physics, such as dark-matter annihilation~\cite{Cholis:2013psa}. 
Pulsars are the rapidly rotating remnant cores of massive stars that have collapsed and exploded leaving behind a highly magnetized neutron star. 
Multiwavelength observations have shown that pulsars primarily accelerate electrons and produce electron and positron pairs via interactions in their magnetospheres. 
These particles, accelerated by powerful magnetic fields generated by the spinning pulsar, interact with the interstellar medium and are advected into pulsar wind nebulae (PWNe)~\cite{Slane:2017bje}.
There is a known population of pulsars whose peak spectral energy distributions (SEDs) lie between 300 keV and 100 MeV~\cite{2015MNRAS.449.3827K}. 
Their signals are likely dominated by pair synchrotron radiation~\cite{Harding:2017ypk}, and they may possess very different leptonic densities relative to the population of pulsars detected by Fermi/LAT in GeV gamma rays~\cite{Fermi-LAT:2013svs}. 
\edit{Based on extrapolating the best fit spectra from the Fermi/LAT catalog 4FGL DR2~\cite{Fermi-LAT:2019yla, Ballet:2020hze} and the 2nd Fermi/LAT catalog of gamma-ray pulsars~\cite{Fermi-LAT:2013svs},}
AMEGO-X will detect more than 15 medium-energy gamma-ray peaked pulsars (mePSRs)~\cite{2015MNRAS.449.3827K, Oh:2018wzc}, observe the phase-resolved spectra of at least 5 mePSRs, and deliver 100 keV - 1 MeV polarization measurements for three pulsars (B1509-58, Vela, the Crab), testing whether synchrotron is the main emission mechanism at MeV energies. 
AMEGO-X will constrain the location of the emission region and the pair multiplicity (i.e., number of electrons and positrons produced), thereby measuring the contribution of different pulsar populations to the CR positron excess.

Pulsars provide an instantaneous snapshot of the number of e$^+$e$^-$ pair produced, while the surrounding PWNe provide a long-term average (tens of thousands of years) of the pairs over the life-time of the pulsar. 
The pulsar relativistic pairs are further accelerated at a wind termination shock at the inner nebula boundary~\cite{1984ApJ...283..710K}. 
The energy of the pairs is regulated by synchrotron radiation losses in turbulent magnetic fields near the shock. 
This sets a natural scale of $\sim$150 MeV for the peak of the SEDs (independent of the field strength~\cite{1996ApJ...457..253D}). 
There PWNe acts like a calorimeter that represents the accumulation of pairs over the history of the pulsar and provides a measure of the net pair energy output. 
AMEGO-X will measure the e$^+$e$^-$ content of more than 10 PWNe, and therefore their contribution to the CR positron excess~\cite{2019BAAS...51c.183D}.
It will also search for counterparts to the extended gamma-ray halos around middle-aged pulsars found by air shower gamma-ray experiments~\cite{HAWC:2017kbo, LHAASO:2021crt}. 

\begin{figure}[ht]
\begin{center}
\includegraphics[width=0.75\textwidth]{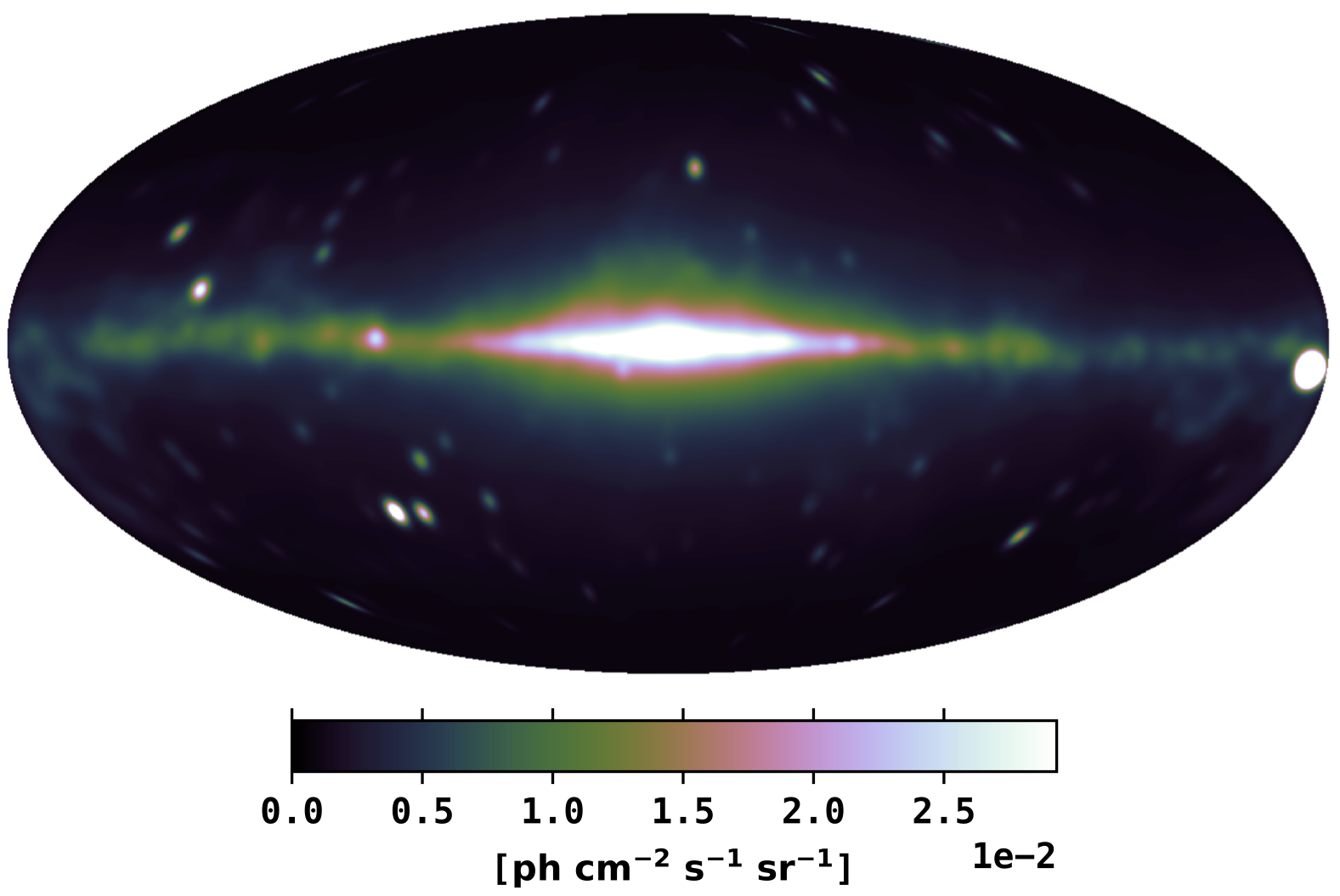} 
\end{center}
\caption{Simulated AMEGO-X 3\,yr all-sky map in the $1-30$ MeV energy range. The map contains Galactic diffuse emission continuum emission calculated with GALPROP~\cite{Moskalenko:1997gh}, as well as individual sources extrapolated from the Fermi/LAT 4FGL-DR2 catalog~\cite{Ballet:2020hze} into the AMEGO-X energy range. Blazars dominate the population ($>$85\%) followed by pulsars. The Galactic diffuse in this energy range primarily includes emission from inverse Compton and Bremsstrahlung (with contribution from $\pi^0$ at higher energy). The map is convolved with a 2D Gaussian kernel to account for the angular resolution of the instrument. We have verified that the Crab flux is in agreement with measurements~\cite{2020ApJ...899..131J}.}
\label{fig:SFO_skymap}
\end{figure}

\subsection{The keV to GeV Gamma-ray Sky}

With the order of magnitude increase in sensitivity in this energy band and based on extrapolations from the Fermi/LAT 10-year source catalog~\cite{Fermi-LAT:2019yla} and Swift/BAT X-ray catalog~\cite{Oh:2018wzc}, AMEGO-X will detect many additional medium-energy gamma-ray--producing sources during normal mission operations. AMEGO-X will deliver science results of significant interest for the astrophysical community and a multi-year catalog of the full medium-energy gamma-ray sky. A simulated sky map is shown in Figure~\ref{fig:SFO_skymap} covering the energy range $1-30$ MeV. The map includes emission from gamma-ray binaries (including accreting black holes in our galaxy), Galactic diffuse continuum emission, and high-redshift blazars~\cite{2019astro2020T.398G, Wadiasingh:2019jdr, DeColle:2019wzp, 2009ApJ...699..603A, 2021arXiv211110362N}. The Galactic diffuse emission is calculated with GALPROP~\cite{Moskalenko:1997gh} and the individual sources are modeled based on an extrapolation of the Fermi/LAT 4FGL-DR2 catalog. Additional sources that will be detectable by AMEGO-X (although not shown in the sky map) include long and short GRBs, magnetar bursts and giant flares~\cite{Wadiasingh:2019jdr}, the extragalactic gamma-ray background~\cite{2009ApJ...699..603A}, and possibly jetted tidal disruption events~\cite{DeColle:2019wzp,Murase:2020lnu} and large scale bubbles~\cite{2021arXiv211110362N}.

\section{AMEGO-X Gamma-Ray Telescope}
\label{Section:Instrument}

\begin{figure}[tb]
\begin{center}
\includegraphics[width=1.0\textwidth]{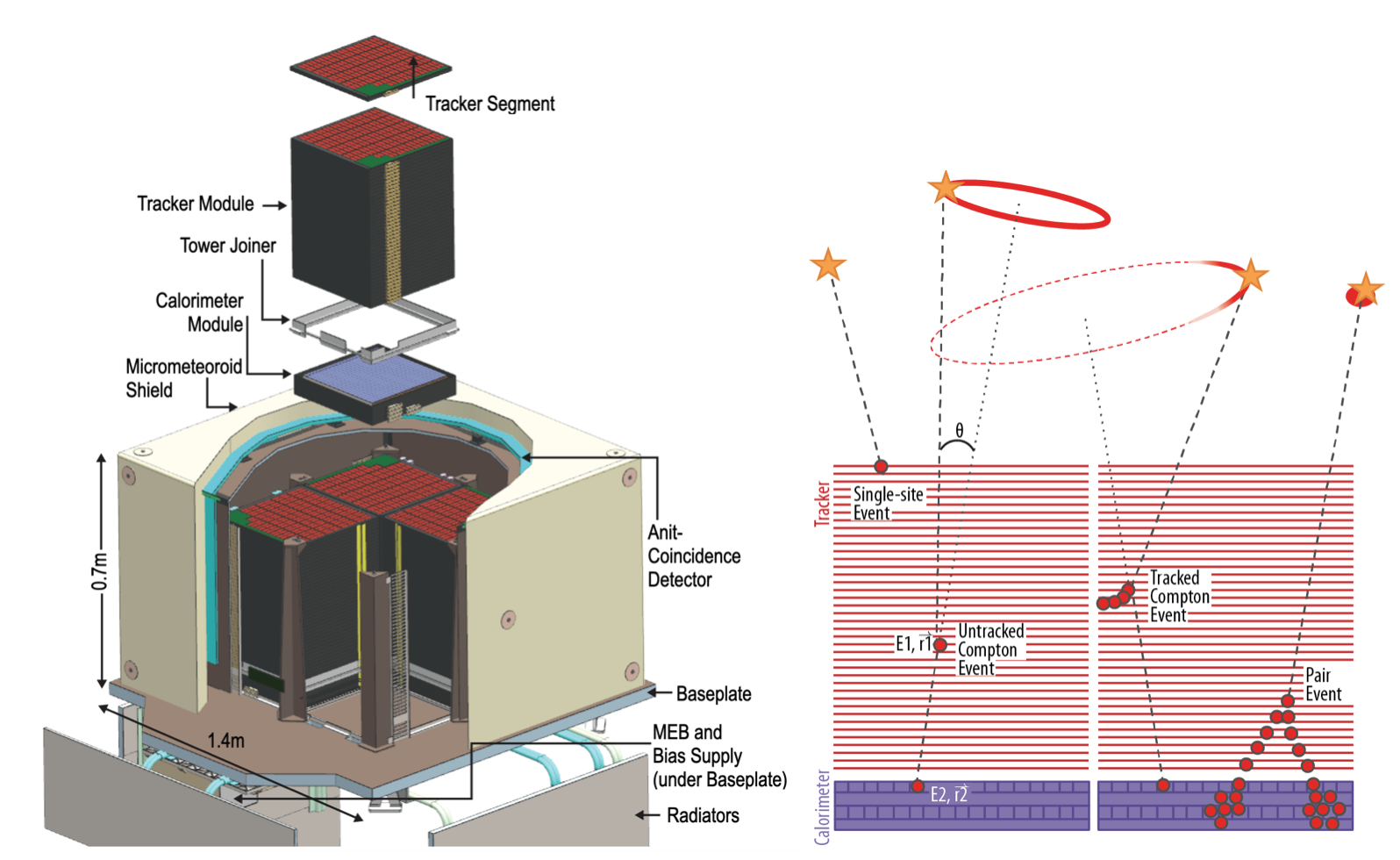} 
\end{center}
\caption{(An exploded view of the Gamma-Ray Telescope on board AMEGO-X (left). The Tracker (red) and Calorimeter (purple) together characterize gamma-rays with three distinct detection techniques (right). Single-site events increase the sensitivity and low energy response ($<$100~keV) for transients only. Untracked and Tracked Compton events provide imaging $<$10~MeV, where the energy ($E$) and position ($\vec{r}$) of interactions are used to determine the initial Compton scatter angle ($\theta$). Pair events enable imaging $>$10~MeV using the same detection techniques as Fermi/LAT.}
\label{fig:AMEGO_Instrument}
\end{figure}

The Gamma-Ray Telescope (GRT) is the AMEGO-X mission’s sole instrument. 
It is a wide-field survey instrument designed to discover and characterize gamma-ray emission from multimessenger sources using imaging, spectroscopy, and polarization.
The GRT is composed of two detector subsystems, the Gamma-Ray Detector (GRD) and the Anti-Coincidence Detector (ACD), which are protected by a Micro-Meteoroid Shield (MMS) (Figure~\ref{fig:AMEGO_Instrument}). 
The GRD consists of a Tracker, with 40 layers of silicon complementary metal-oxide-semiconductor (CMOS) monolithic Active Pixel Sensors (APS), and a Calorimeter, with four layers of Cesium Iodide (CsI) scintillator bars. 
Together, they characterize gamma rays from 100 keV (25 keV for transients) to 1 GeV~\cite{Fleischhack:2021mhc}.
The GRT baseline capabilities are summarized in Table~\ref{tab:params}.

The MeV range covers three different photon-matter interactions that dictate three detection techniques for GRT, as shown in Figure~\ref{fig:AMEGO_Instrument}.
Between $\sim$100 keV and $\sim$10 MeV, photons predominantly Compton scatter. 
The measured position and energy of a Compton scattering interaction and subsequent absorption kinematically and geometrically constrain the initial direction of the primary gamma ray to a circle in the sky\cite{Boggs:2000zk}. 
Such an event is referred to as an untracked Compton event. 
Compton scattering is inherently polarization sensitive, and a linearly polarized source generates a sinusoidal scattering angle distribution in the instrument~\cite{1997SSRv...82..309L}. 
If the direction of the first Compton-scattered electron is measured in the Tracker, this additional kinematic information constrains the photon direction to an arc and these tracked Compton events allow for improved background rejection~\cite{2004NewA....9..127A}.
High-energy gamma rays ($>$10 MeV) convert to an electron-positron pair, which in turn is detected through ionization tracks in the instrument. 
The direction of the incoming photon is determined from the positions of the interactions in the Tracker and the total energy is determined by the electromagnetic shower(s) detected in the Calorimeter~\cite{2009ApJ...697.1071A}.

At energies below the Compton regime ($<$100 keV), photons predominantly undergo photoelectric absorption in a single pixel in the Tracker.
While these single-site events  have no imaging capability, they can be used to localize transient sources using the aggregate signal\cite{Martinez-Castellanos:2021bbl}.
For enhanced low-energy sensitivity to GRBs, AMEGO-X enables short duration ($<$100 s) readout of single-site events to measure emission down to the Tracker threshold of 25 keV~\cite{Martinez-Castellanos:2021bbl}.

\begin{table}[tb]
\centering
\caption{The Gamma-Ray Telescope baseline capabilities.}
\begin{tabular}{|l|c|}
\hline
 Parameter & \\ \hline \hline
 Energy Range & 25 keV -- 1 GeV  \\ 
 Energy Resolution & 5\% FWHM at 1 MeV, 17\% (68\% containment half width) at 100 MeV \\  
 Point Spread Function & 4$^{\circ}$ FWHM at 1 MeV, 3$^{\circ}$ (68\% containment) at 100 MeV  \\
 Localization Accuracy & transient: 1$^{\circ}$ (90\% CL radius), persistent: 0.6$^{\circ}$ (90\% CL radius)  \\ 
 Effective Area & 1200 cm$^2$ at 100 keV, 500 cm$^2$ at 1 MeV, 400 cm$^2$ at 100 MeV \\  
 Field of View & 2$\pi$ sr ($<$10 MeV), 2.5 sr ($>$10 MeV) \\ \hline
\end{tabular}
\label{tab:params}
\end{table}

\subsection{Gamma-ray Detector}

The GRD is the primary GRT science subsystem. 
It consists of four identical Detection Towers (Figure~\ref{fig:AMEGO_Instrument}), each with a Tracker and Calorimeter Module, a dedicated low-voltage power supply (LVPS), and Digital Input and Output (I/O) board. 
Although each Detection Tower operates independently, signals are combined in the Main Electronics Box (MEB) such that events are reconstructed using data from the full GRD. 
The Towers’ data acquisition (DAQ) electronics and thermal management hardware are positioned along the sides of the GRD to reduce the amount of passive material within the sensitive instrument volume.

\subsubsection{Pixelated Silicon Tracker}

\begin{figure}[t]
\begin{center}
\includegraphics[width=1.0\textwidth]{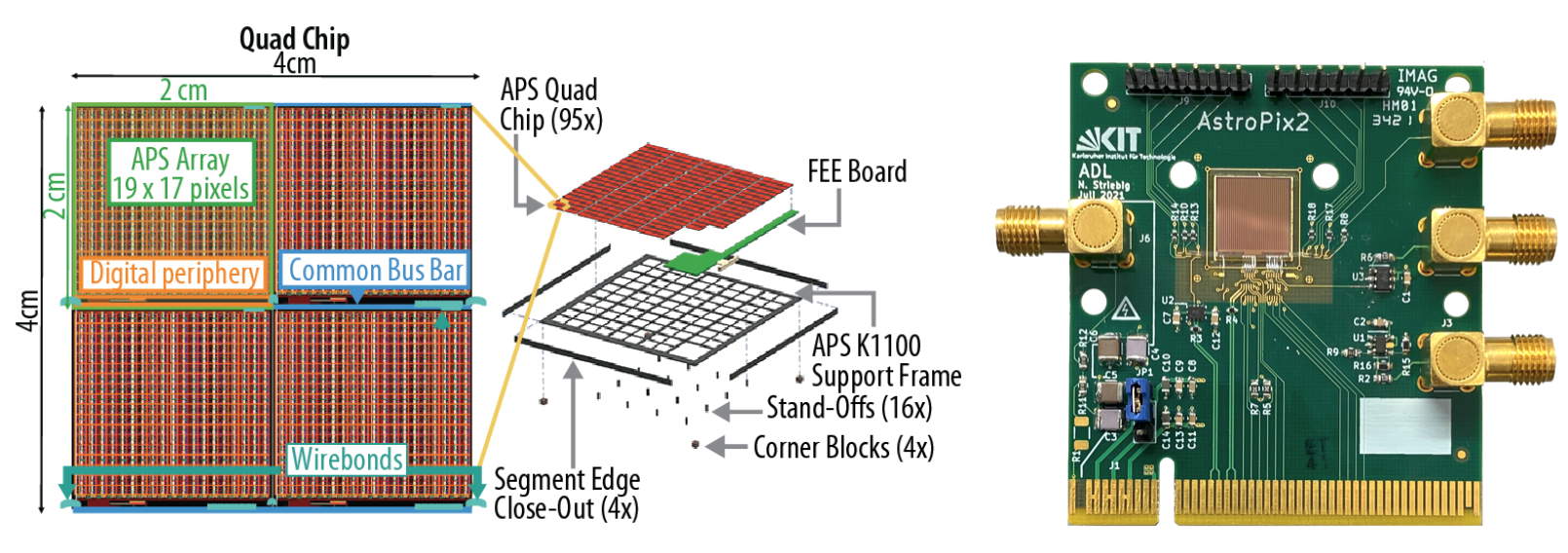} 
\end{center}
        \caption{The AMEGO-X Tracker relies on low-noise, low power CMOS monolithic APS detectors (left). Each Quad Chip consists of four identical APS arrays of $19 \times 17$ pixels, with the digital periphery located at the bottom of each APS array. Each Tracker \edit{Segment} consists of 95 Quad Chips with a common bus bar for power distribution and wirebonds utilizing SPI readout. The Quad Chips are supported with a K1100 carbon fiber frame. The prototype AstroPix$\_$v2 detector, a $1 \times 1$~cm chip, has been fabricated and tested on the bench and in heavy ion beam tests (right). 
        }
        \label{fig:SFO_APS}
\end{figure}

The GRD Tracker’s main functionality is to measure the energy and position of gamma-ray and charged-particle interactions with high precision. 
Each of the four Tracker Modules consist of 40 identical stacked Tracker Segments (45$\times$45~cm$^2$) of silicon APS detectors, separated by 1.5~cm. 
The Tracker Segments each contain 95 Quad Chips (Figure~\ref{fig:SFO_APS}), which consist of four identical APS arrays cut out from a single silicon wafer. 
The AMEGO-X APS chip, AstroPix, is a 2$\times$2~cm$^2$ array of 19$\times$17 pixels measuring 1$\times$1~mm$^2$.
Each pixel contains a charge-sensitive preamplifier and comparator, where the active circuitry within the pixel results $<$1\% loss in charge-collection volume. 
The APSs are 0.5 mm thick and operate at full depletion.

The major strength of the APS detectors is low noise, which is achieved through integration of the readout electronics within the detecting material. 
The AMEGO-X APS performance parameters have been determined through measurements of ATLASPix (designed for the ATLAS experiment at CERN)~\cite{Brewer:2021mbe} and the first two prototype versions of AstroPix (Figure~\ref{fig:SFO_APS}), in addition to simulations from the designers at Karlsruhe Institute of Technology (KIT). 
AMEGO-X leverages more than 10 years of development in CMOS monolithic active pixel silicon detectors from ground-based particle physics experiments~\cite{Peric:2007zz, 2019NIMPA.924...99P, Schoning:2020zed, Peric:2021bcu}. 
By optimizing these APSs for space applications~\cite{10.1117/12.2562327, Brewer:2021mbe} the detectors enable observations at lower photon energies, achieve an overall increase in sensitivity, and are simpler to integrate compared to previous silicon detector technologies.

The GRT design uses minimal passive material and carbon fiber reinforced polymers to reduce photon attenuation and backgrounds from activation.
In this regard, the Tracker is designed with high thermal conductivity K1100/Cyanate Ester (CE) for heat extraction from the APS detectors. The APS Support Frame (Figure~\ref{fig:SFO_APS}) is CNC cut K1100/CE laminate bonded to M55J/CE perimeter closeouts and stand-offs for additional stiffness.

\subsubsection{Hodoscopic CsI Calorimeter}

\begin{figure}[t]
\begin{center}
\includegraphics[width=1.0\textwidth]{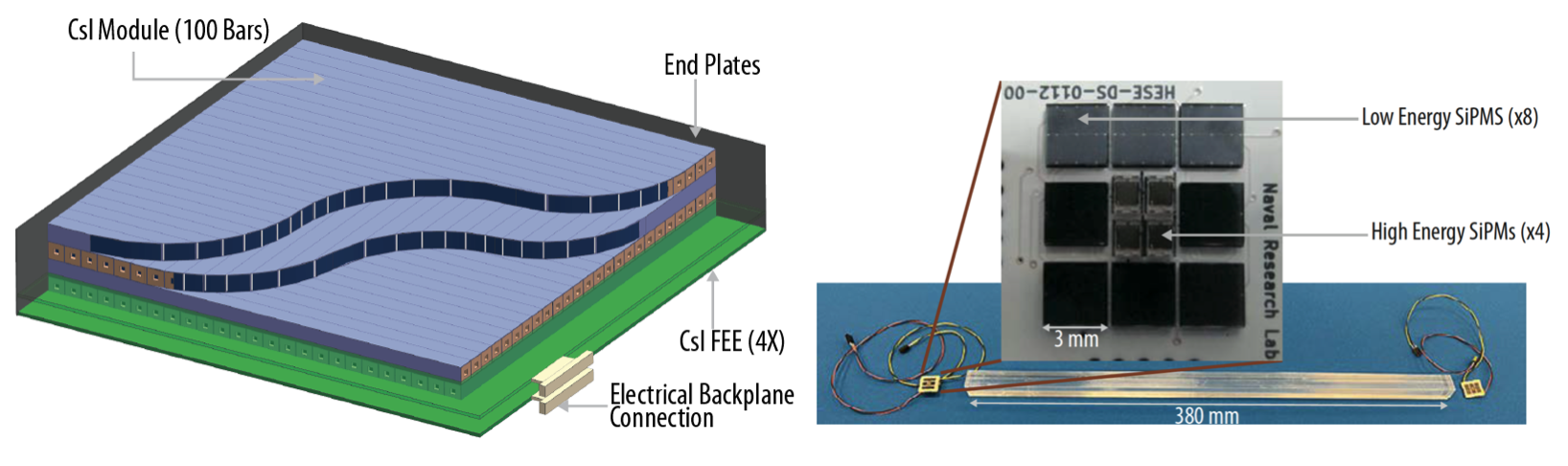} 
\end{center}   
        \caption{The GRT hodoscopic arrangement of four Calorimeter layers enables an accurate measurement of gamma-ray energies through the profile of electromagnetic showers (left). The dual-gain SiPM readout at the ends of each CsI bar has been demonstrated in the lab and enables low-noise readout across a large dynamic range (right). 
        }
        \label{fig:SFO_CAL}
\end{figure}

The main functionality of the Calorimeter is to measure the position and energy of Compton-scattered photons and the electromagnetic showers produced from electron and positron pairs over a broad energy range. 
Situated directly below the Tracker subsystem, the Calorimeter is composed of four layers of thallium-doped cesium iodide (CsI:Tl) bars, hodoscopically arranged. One of the four Calorimeter Modules is shown in Figure~\ref{fig:SFO_CAL}.

Each layer consists of 25 bars, each with a dimension of 1.5$\times$1.5$\times$38~cm$^3$. 
The bars are wrapped in reflective material to pipe scintillation photons to each end, where readout occurs via an array of silicon photomultipliers (SiPMs). 
To achieve a large dynamic range in a single Calorimeter bar and to mitigate the effects of saturation, 
a mixture of small and large ONSemi SiPMs are used to cover two gain ranges~\cite{shy2021}, as shown in Figure~\ref{fig:SFO_CAL}. 
Each end of each CsI bar is read out with a low energy SiPM array, which is a sum of eight 3$\times$3~mm$^2$ SiPMs, and a high energy array, which is a sum of four 1$\times$1~mm$^2$ SiPMs.
\edit{Although SiPMs are sensitive to damage from the orbital radiation environment, work done at NRL has demonstrated that the damage effects can be successfully mitigated by proper instrument configuration without affecting instrument performance~\cite{2017SPIE10397E..0BM, Mitchell:2020dii, 2019arXiv190711364M}.} 
The position of the interaction along the bar is determined from the relative amplitude of signals on each end.

The AMEGO-X Calorimeter utilizes a design based on Fermi/LAT~\cite{2009ApJ...697.1071A}. The Calorimeter team at Naval Research Laboratory (NRL) designed, developed, assembled, tested, and currently operates the Fermi/LAT CsI Calorimeter. Furthermore, the team has built and demonstrated the performance of a prototype AMEGO-X GRT Calorimeter with SiPM readout in gamma-ray beam tests~\cite{8824629, shy2021}.

\subsection{Anti-coincidence Detector}

\begin{figure}[tb]
    \centering
    \includegraphics[width = 0.5\textwidth]{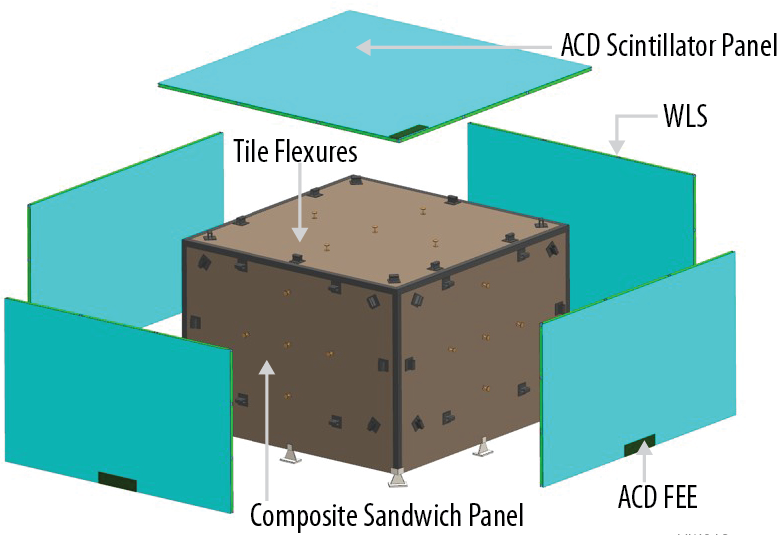}
    \caption{The AMEGO-X ACD is a simple, non-segmented version of the Fermi/LAT design. The ACD consists of 5 scintillation panels read out through wave-length shifting (WLS) bars with SiPM arrays on each end. The ACD surrounds the GRD to veto incident charged particles, reducing the background rate during flight.}
    \label{fig:ACD}
\end{figure}

AMEGO-X uses a dedicated Anti-Coincidence Detector (ACD) to reduce the significant cosmic-ray background. 
The ACD comprises five plastic scintillator panels that surround the GRD to enable vetos associated with incident charged particles while being transparent to gamma rays (Figure~\ref{fig:ACD}). 
Each ACD panel has three wavelength-shifting bars (WLS) on each of its four edges. 
Scintillation light entering the WLS bars is transmitted to the ends, where signals are measured by arrays of 6$\times$6 mm$^2$ ONSemi SiPMs. 
Each WLS is read out independently to recover the energy deposited in each ACD panel. 
The ACD has an energy threshold of 200 keV, which is well below the minimum ionizing particle (MIP) average energy deposition of 2.5 MeV.
The ACD interfaces directly with the MEB, which provides I/O and FEE power.

\section{Instrument Performance}
\label{sec:performance}

For Compton and pair telescopes, much of the performance relies on accurate event reconstruction and background rates, and therefore simulations were performed with the state-of-the-art MEGAlib analysis package~\cite{2006NewAR..50..629Z}, which is built around Geant4\cite{AGOSTINELLI2003250}. 
A detailed mass model that reproduces the active and passive material is simulated with the detector performance to determine interactions of photons and particles and the resulting measured signals. 

\subsection{Background Model}

Background radiation is one of the dominant factors that can limit the sensitivity of an MeV telescope, and therefore detailed background simulations are necessary for accurate performance predictions.
The background models used in MEGAlib are based on measurements by COMPTEL~\cite{Gruber:1999yr}, INTEGRAL/SPI~\cite{2003A&A...411L.113W}, Fermi/LAT~\cite{Mizuno:2004jb}, and NuSTAR~\cite{10.1093/mnras/stab209}, and cosmic-ray population measurements, with an assumed AMEGO-X orbit of 575~km and an inclination of 6$^{\circ}$. 
The models and components (including prompt and delayed emission from cosmic-rays, extragalactic diffuse, and Albedo emission) are \edit{orbit averaged and are} described \edit{in more detail} in Ref.~\cite{Cumani:2019ryv}.

\begin{figure}[t]
    \centering
    \includegraphics[width=0.6\textwidth]{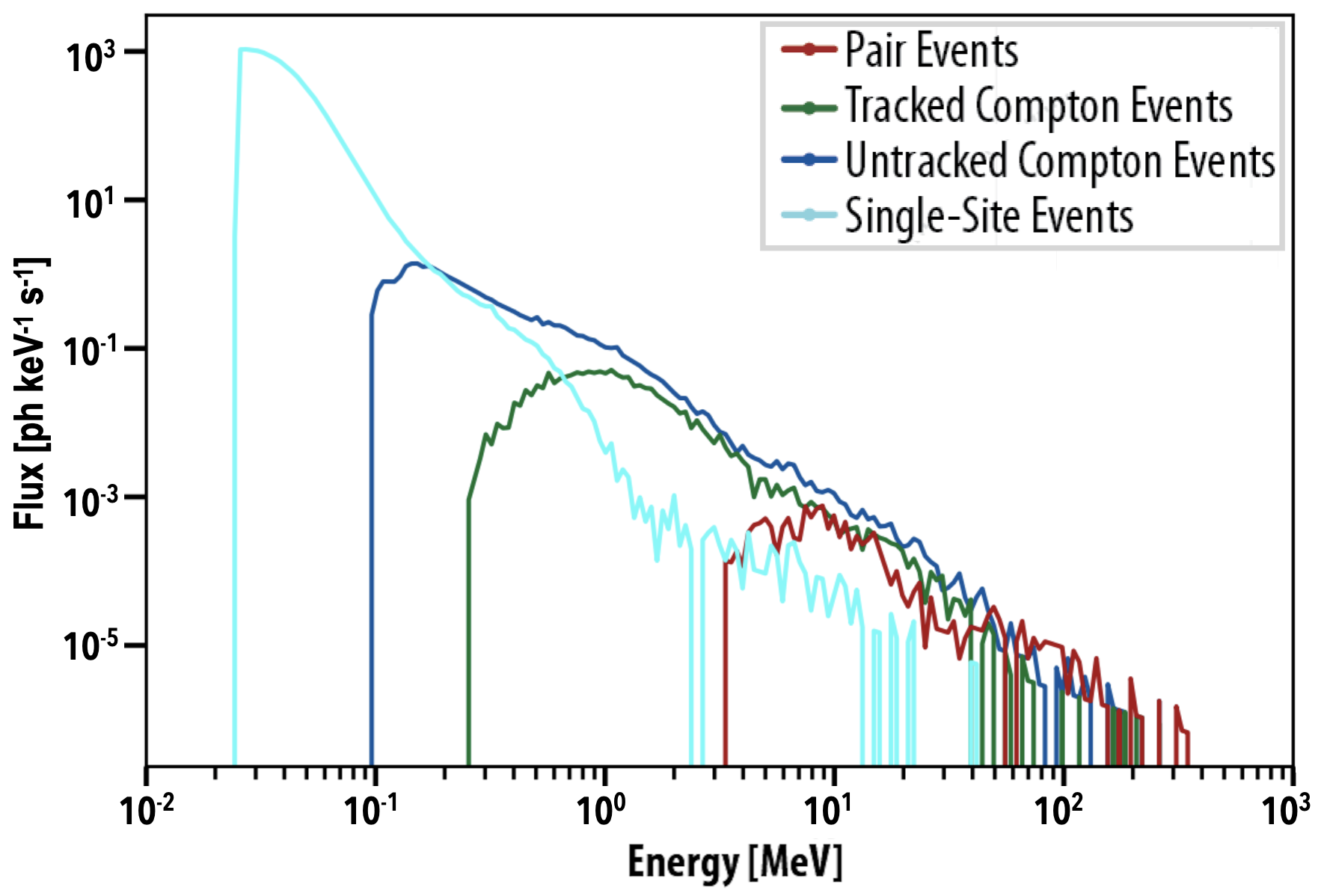}
    \caption{The expected measured rates for each event type from the MEGAlib background models. The event classification is performed in MEGAlib, and the dominant background at lower energies is cosmic and atmospheric photons, while at higher energies in the Compton regime, the delayed emission of activated material in the instrument is the dominant source of background. The majority of background events $>1$~MeV are classified as untracked Compton events because this is the most nonrestrictive event class.}
    \label{fig:bkg}
\end{figure}

The measured signals from charged cosmic rays is a significant background to gamma-ray observations, but almost all of these events can be vetoed by the ACD. Another background-rejection capability is through event reconstruction; only events which have a valid Compton sequence or pair-conversion tracks will be identified as ``good'' events for higher-level analysis.
Figure~\ref{fig:bkg} shows the resulting measured background spectrum after the ACD veto and event reconstruction, separated for each event type as classified by MEGAlib \cite{Martinez-Castellanos:2021bbl}.
The most significant background rate is measured at low energies as single-site events, which correspond to a single triggered pixel in the Tracker. This measured flux is dominated by cosmic and atmospheric photons, and without imaging capabilities to separate any source emission from background, single-site events can only be used for transient detection when the source rate is high. 
Most background events which have more than one trigger in the GRT, but are not vetoed by the ACD and do not leave clear straight charged-particle tracks in the Tracker, are classified as Untracked Compton events, since that is the most inclusive event category in MEGAlib.
The dominant background component in the Compton regime is activation of the passive and active instrument material by cosmic rays, where the delayed emission of radiation cannot by vetoed by the ACD.
\edit{In the Pair regime, the dominant background is albedo photons~\cite{Cumani:2019ryv}, which also are not vetoed by the ACD.} 

Further background rejection can be achieved through fine-tuned event selections depending on the observation, such as the total photon energy, pair opening angle, Compton scatter angle, or distance between interactions.

\subsection{Angular Resolution and Effective Area}

\begin{figure}[t]
\begin{center}
\includegraphics[width=1.0\textwidth]{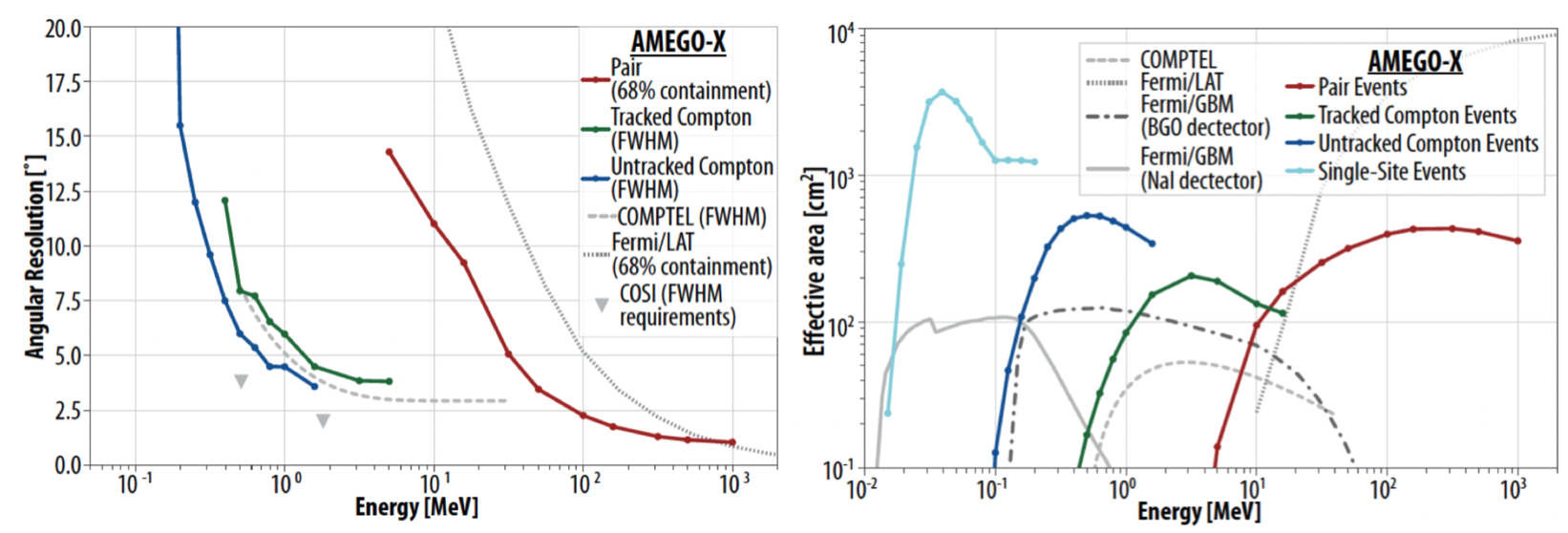} 
\end{center}
\caption{The projected angular resolution (left) and effective area (right) are determined through on-axis, mono-energetic, point source simulations. In addition to background rate, these parameters are the main contribution to the sensitivity of the instrument.}
\label{fig:PFO_aeff_ares_eres}
\end{figure}

Monoenergetic point source simulations are performed in MEGAlib to determine the energy resolution, effective area, and angular resolution of AMEGO-X. 
Figure~\ref{fig:PFO_aeff_ares_eres} shows the measured angular resolution and effective area as a function of energy for each event classification from these simulations. 
The angular resolution in the Compton regime is defined as the FWHM of the angular resolution measure (ARM) histogram, which is a projection of the point spread function in one dimension, as is standard for Compton telescopes. 
In the pair regime, the angular resolution is defined as the 68$\%$ containment radius, following the standard Fermi/LAT definition. 
There is no angular resolution for the single-site events since there is no imaging capabilities from these interactions.

The predicted angular resolution is $4^{\circ}$ at 1 MeV and $\edit{2.5}^{\circ}$ at 100 MeV. 
Measuring the track of a Compton-scattered electron does not improve the angular resolution in the Compton regime as the uncertainty in the scattering angle is often large. The angular resolution for Tracked Compton events are in fact slightly worse than those for Untracked Compton events due to the inherent selection of events which have larger scattering angles. 
Also shown in Figure~\ref{fig:PFO_aeff_ares_eres} are the measured angular resolution of COMPTEL, Fermi/LAT, and the requirements for COSI. The angular resolution of AMEGO-X is better than Fermi/LAT since the LAT's resolution is limited by multiple scattering in the tungsten conversion foils within the tracker. 

The simulated effective area, shown in Fig.~\ref{fig:PFO_aeff_ares_eres}, is a measure of the detection efficiency of the telescope. It is defined as the required area of an ideal detector, i.e. 100\% efficient, to detect an equivalent number of photons. 
It is calculated here as the number of valid events classified by MEGAlib divided by the initial number of simulated photons, and scaled by the area of the mass model's surrounding sphere. 
The expected effective area is 1200~cm$^{2}$ at 100~keV, 500~cm$^2$ at 1~MeV, and 400~cm$^2$ at 100~MeV. Fig.~\ref{fig:PFO_aeff_ares_eres} also shows the effective area for a single Fermi/GBM BGO and NaI detector, as well as COMPTEL and Fermi/LAT.

\subsubsection{Continuum and Transient Sensitivity}

\begin{figure}[t]
\begin{center}
\includegraphics[width=0.6\textwidth]{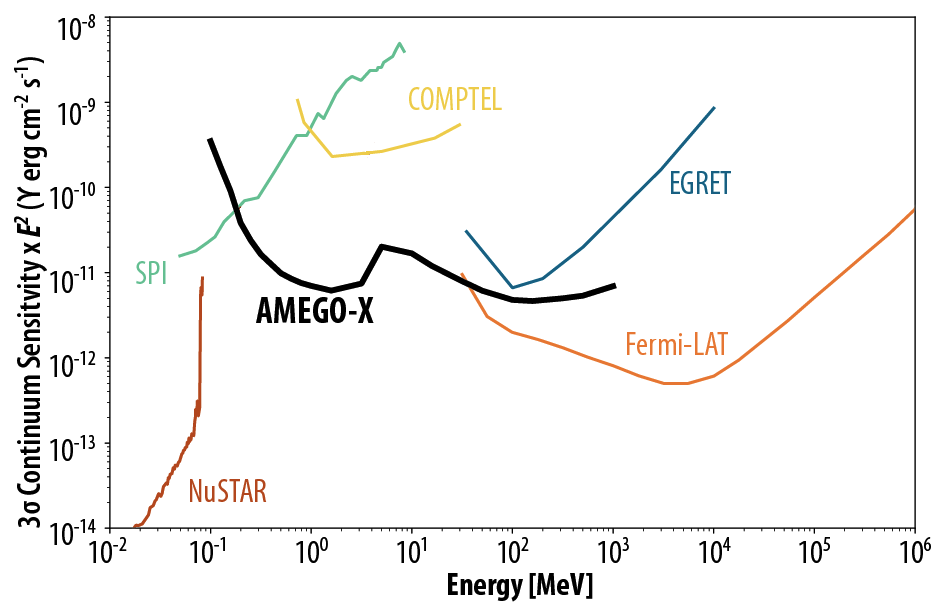}
\end{center}
\caption{The AMEGO-X continuum sensitivity \edit{using Untracked Compton, Tracked Compton, and Pair events} is shown for the three year mission compared to the sensitivity of past and present missions. The effective area is conservatively taken to be for sources 37$^{\circ}$ off-axis, and the observation time is estimated to be 20\% of the full mission (T$_{\mathrm{Obs}}$ = 3 yr $\times$ 0.2).}
\label{fig:d7}
\end{figure}

The sensitivity of AMEGO-X is determined for steady-state (i.e. continuum) sources and transient sources based on the above MEGAlib simulated performance parameters.
From the measured effective area, angular resolution, and energy resolution, the continuum sensitivity can be calculated as the minimum detectable source flux: 
\[ F_\mathrm{min}(E)=\frac{n^2+n\sqrt{(n^2+4N_b )}}{2 A_{\mathrm{eff}}  T_{\mathrm{Obs}} } \]
where $n$ is the required detection significance (3$\sigma$) per energy band, $N_b$ is the number of background counts, $A_{\mathrm{eff}}$ is the effective area, and $T_\mathrm{Obs}$ is the observation time. 
The 3-year AMEGO-X sensitivity is shown in Figure~\ref{fig:d7}.
To account for the survey-mode observation, the effective area is conservatively taken from simulations of sources at 37$^{\circ}$ off-axis, and the observation time is estimated to be 20\% of the full mission ($T_{\mathrm{Obs}}$ = 3 yr $\times$ 0.2). 

The continuum sensitivity is calculated for Untracked Compton, Tracked Compton, and Pair events separately. 
When there are significant contributions from two event types, for example Pair events at 10~MeV have a similar effective area to Tracked events at this energy,  the sensitivities are combined: 
\[ F_{\mathrm{combined}}= F_{\mathrm{tracked}} \sqrt{\frac{F_{\mathrm{pair}}}{F_{\mathrm{tracked}}+F_{\mathrm{pair}}}} \]
The continuum sensitivity is $3.2\times10^{-12}$~erg/cm$^2$/s at both 1~MeV and 100~MeV. 
The regime where Compton and Pair events overlap, around 10~MeV, has worse sensitivity due to the limited classification capability currently implemented in MEGAlib. 
Recent developments in event identification and reconstruction with neural networks have shown dramatic improvements in sensitivity~\cite{2020AAS...23537221Z}. Additionally, the projected performance in the pair regime ($>$10 MeV) is expected to improve, as MEGAlib does not currently include reconstruction and shower profiling techniques developed for Fermi/LAT~\cite{2012ApJS..203....4A}.

The transient sensitivity is determined from the measured signal-to-noise ratio from simulations of canonical GRBs. It is defined for each event type (single site and Compton): 
\[ SN=\frac{S}{\sqrt{S+B}} \] 
where $S$ is the number of source counts detected (from GRB source simulations) and $B$ is the number of background counts (from background simulations). 
The minimum detectable flux at which the combined SN ratios $F_{\mathrm{min}} = \sqrt{SN_{\mathrm{single~site}}^2+SN_{\mathrm{Compton}}^2}$
$\geq\,6.5\,\sigma$ and $SN_{\rm Compton}\geq$4.5\,$\sigma$ is the transient sensitivity. The AMEGO-X transient sensitivity is $0.5$~$\gamma$/cm$^2$/s between 25 \edit{keV}--1~MeV for 1 second.
The derived GRB rates account for the dependence of the effective area with off-axis angle, and the transient sensitivity is described \edit{in Ref.}~\cite{Martinez-Castellanos:2021bbl}.

\section{AMEGO-X Mission Implementation}
\subsection{Spacecraft}

\begin{figure}[tb]
\begin{center}
\includegraphics[width=0.7\textwidth]{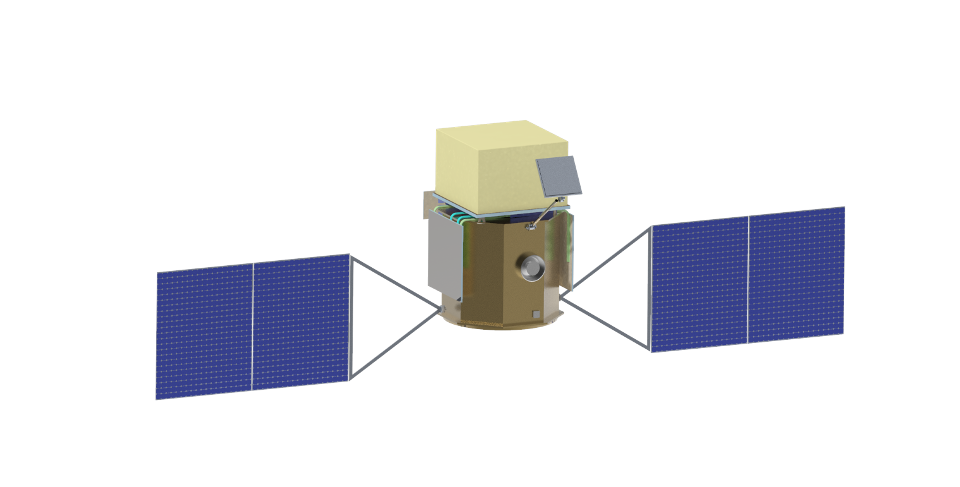}
\end{center}
\caption{The Gamma-ray Telescope (yellow box) is shown on top of a base plate attached to the spacecraft bus. The pair of single-axis gimbaled solar arrays from end to end are approximately 11 m wide. This figure illustrates the deployed high-gain antenna, and one of the two low-gain antennae attached to the spacecraft bus. The other low-gain antenna is on the back side. Radiators (also gray) are located on each side of the spacecraft bus other than the side with the high-gain antenna. }
\label{fig:f5}
\end{figure}

The AMEGO-X flight system consists of the GRT instrument integrated onto a spacecraft bus which leverages Lockheed Martin Space's (LMS) standard subsystem architectures (Figure~\ref{fig:f5}). 
These include structure, mechanisms, power, attitude determination and control (ADCS), C\&DH, and flight software (FSW) from cost-capped planetary missions dating back $>$15 years, and hardware and software from LMS spacecraft including MAVEN, Lucy, IRIS, OSIRIS-REx, and Juno. 

\subsection{Observation plan}

\begin{figure}[tb]
\begin{center}
\includegraphics[width=0.5\textwidth]{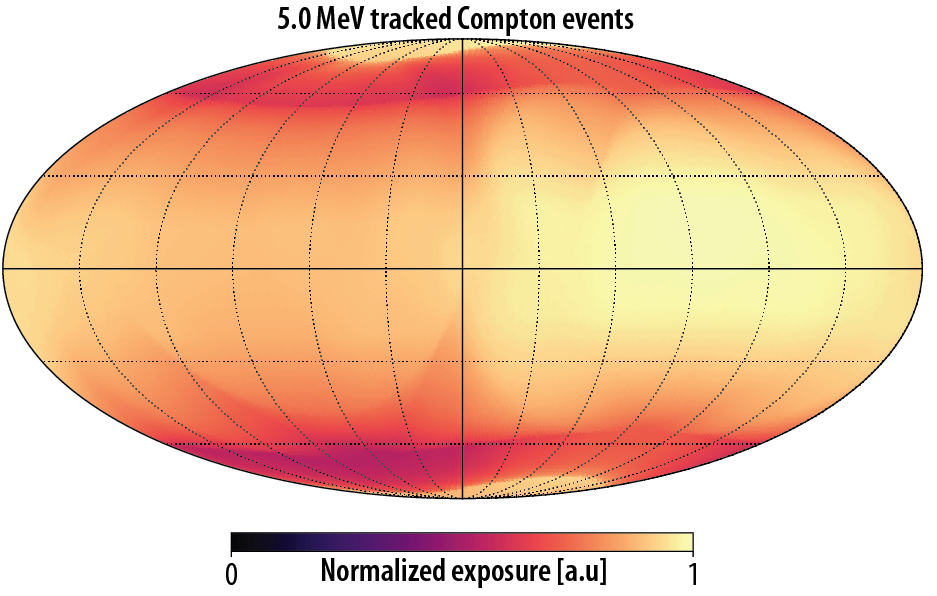} 
\end{center}
\caption{The normalized exposure map for 5 MeV gamma rays (as an illustration) assuming low-Earth orbit (600 km) with a low inclination (5$^{\circ}$). Rocking alternately $\pm$30$^{\circ}$ from zenith every orbit enables the GRT to uniformly observe nearly the entire sky every two orbits ($\sim$3 hours). Over the course of 24 hours, the GRT survey covers the entire sky. }
\label{fig:e8f1}
\end{figure}

AMEGO-X has been proposed to launch into Low Earth Orbit (LEO) in 2028 to start a 3-year baseline science campaign with the potential for an extended mission. 
Science campaign operations are straight-forward, consisting of scanning the sky with the GRT instrument 30$^{\circ}$ North or South of zenith every other orbit, providing a nearly all-sky view every two orbits (Figure~\ref{fig:e8f1}) . 
The baseline AMEGO-X LEO orbit (Figure~\ref{fig:e8f1}) is circular with a 600 km altitude and inclination of 5$^{\circ}$. 
This inclination provides the low background radiation environment needed to achieve the required GRT sensitivity. 

\subsubsection{Science Data Plan and Products}

The AMEGO-X mission performs observations of the full medium energy gamma-ray sky every three hours. 
The data include four different gamma-ray event classes distinguished by the energy of the incoming gamma-ray: Single-Site (photoelectric absorption, used only for fast transients), Compton (Tracked or Untracked), and Pair events.

\begin{figure}[tb]
\begin{center}
\includegraphics[width=0.7\textwidth]{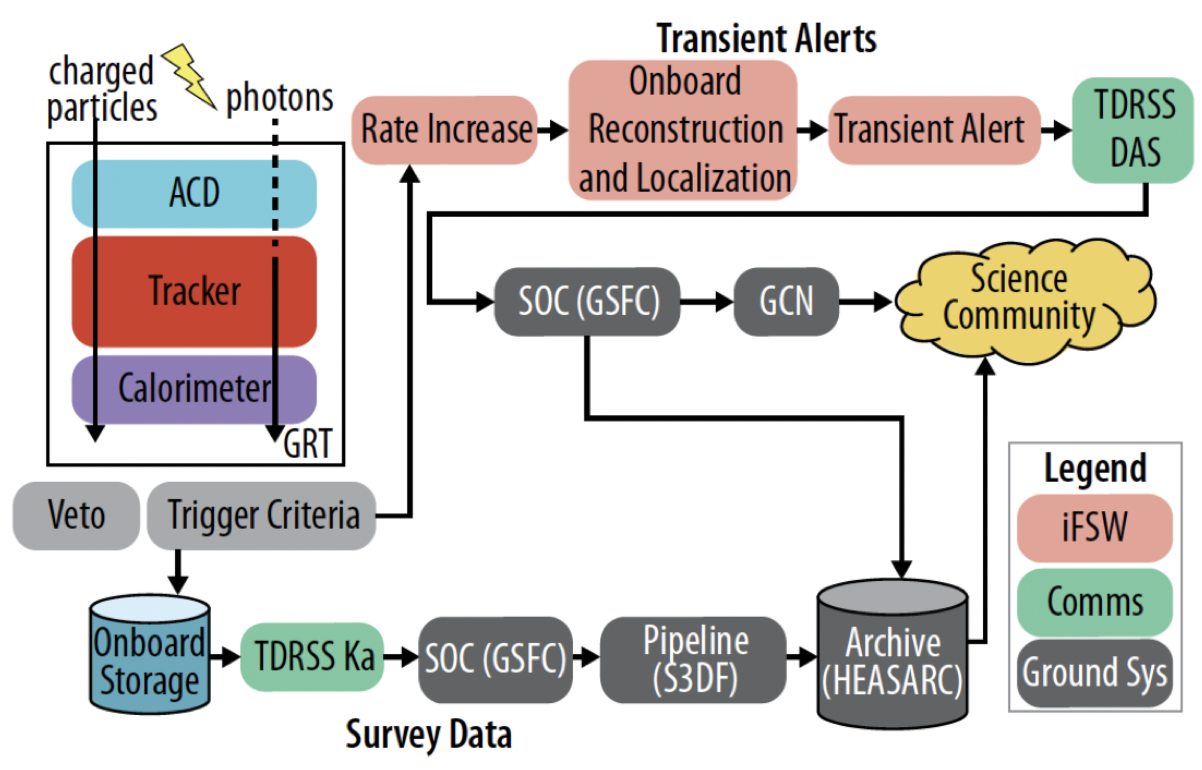} 
\end{center}
\caption{The two main science data products, Transient Alerts and Survey Data, use the same mode of operation and pipeline, but different telemetry paths.
Transient Alerts (90\% in 30 s) use Tracking and Data Relay Satellite System (TDRSS) for rapid alerts to the Gamma-ray Coordinates Network (GCN) for mulitimessenger and multiwavelength followup.
All-sky Survey Data (90\% in 24 hrs) is telemetered via TDRSS Ka band 2-3 times per day for on-ground processing and dissemination to the science community.}
\label{fig:SFO2_j}
\end{figure}

AMEGO-X data will be event-based, where each gamma-ray interaction in the instrument is analyzed separately. 
Images, light curves, polarization analysis, and other science products are generated on the ground as illustrated in Figure~\ref{fig:SFO2_j}. 
For the rapid detection and localization of gamma-ray transients, AMEGO-X uses on-board Transient Alert (TA) Logic based on Fermi/GBM algorithms~\cite{2009ApJ...702..791M, 2012ApJS..199...18P}. 
TAs are identified by the MEB CPU as a significant (6.5$\sigma$) rate increase above the background through a combination of Tracker triggers and events with triggers in Tracker and Calorimeter. 
Based on simulations, TAs are expected 3-5/day and the probability of a false TA detection is $<$1 per year.

The AMEGO-X Data Plan identifies multiple levels of scientific data products that begin as raw binary files downlinked from the spacecraft and end as scientific data products, such as source spectra and light curves. 
Users download photon and housekeeping data from the science archive, along with instrument response functions, diffuse maps, and source lists used to perform analyses on any time, energy, and area scale.
The TA data contains the spacecraft attitude information, a coarsely binned light curve, and localization from on-board event reconstruction. 
After straightforward Science Operations center (SOC) processing, the TA is sent to the Gamma-ray Coordinates Network (GCN) to enable multiwavelength follow-up. 
After the survey data is processed on the ground, a more accurate localization and full light curves will update the initial GCN alert.
The GRT instrument team will provide python-based analysis tools and tutorials to make AMEGO-X data accessible to the scientific community, similar to Fermi/LAT~\cite{2017ICRC...35..824W} and current tools being developed for COSI. 
Photon and spacecraft data will be accessible through a custom photon data server (similar to Fermi/LAT). 
Weekly photon and spacecraft files, all catalogs, and higher-level data products will be available via the High Energy Astrophysics Science Archive Research Center (HEASARC).

\section{Summary}
AMEGO-X will deliver breakthrough discoveries \edit{as a MIDEX class mission} addressing areas highlighted as the highest scientific priority for Explorer-scale missions in the Astro2020 Decadal Survey Report: multimessenger astrophysics and time-domain astronomy.
During its three year mission, it will survey the gamma-ray sky with unprecedented sensitivity in the energy range from 100 keV to 1 GeV. 
The Gamma-Ray Telescope design is well understood with a combination of instrument subsystems that leverage large investments in detector technologies by the Department of Energy, now tailored for space use, while also taking advantage of the extensive flight heritage from Fermi.
Its all-sky coverage enables a sensitivity to transients from milliseconds to years. 
AMEGO-X is complementary to Fermi and COSI with broad MeV continuum, transient and polarization capabilities. 

The AMEGO-X team has experts from Fermi and COSI who have built the instruments, simulation software, data pipelines and data analysis tools. 
AMEGO-X’s spacecraft partner, Lockheed Martin Space (LMS), has a demonstrated track record of successful Explorer-class bus design and operation. 
The AMEGO-X science and instrument teams include members from NASA GSFC, Argonne National Laboratory, the Naval Research Laboratory, and LMS, as well as university science partners who are members of the LIGO Collaboration, the IceCube Collaboration, and the Cherenkov Telescope Array Consortium. 
Together the science team will ensure the maximum return from this unique and groundbreaking mission.


\subsection* {Acknowledgments}
The authors would like to acknowledge generous ongoing support from a number of agencies and institutes that have supported the development of the AMEGO-X mission concept. 
These include NASA Goddard Space Flight Center, the Naval Research Laboratory, and Argonne National  Laboratory. 
The material is based upon work supported by NASA under award number 80GSFC21M0002. 
We also acknowledge the contributions to the design of the AMEGO-X spacecraft by Jonathan Hartley and Lindsay Papsidero from Lockheed Martin Space and the design of the AMEGO-X instrument by the engineering team at Goddard Space Flight Center.


\bibliography{main}   
\bibliographystyle{spiejour}   



\listoftables
\listoffigures

\end{spacing}
\end{document}